# Cubic-$Zn_{1-x}Mn_xTe$ optical modes across the phonon and phonon-polariton regimes

M. Mina,[1] A. Elmahjoubi,[2] T. Ngulezhu,[1,3] S. Thiam,[1] A. V. Postnikov,[1] T. Tite,[1] P. Bourges,[2] F. Bourdarot,[4] N. Biniskos,[4] A. Polian,[5] Y. Le Godec,[5,6] N. Guignot,[6] J.-P. Itié,[6] S. Diliberto,[7] S. Michel,[7] A. Maillard,[8] A. En Naciri,[1] L. Broch,[1] P. Franchetti,[1] K. Strzałkowski[3] and O. Pagès[1,*]

1 Université de Lorraine, LCP-A2MC, F-57000 Metz, France
2 Laboratoire Léon Brillouin, CEA-CNRS, Saclay, France
3 Institute of Physics, Faculty of Physics, Astronomy and Informatics, Nicolaus Copernicus University in Toruń, ul. Grudziądzka 5, 87-100 Toruń, Poland
4 Université Grenoble Alpes, CEA, IRIG, MEM, MDN, F-38000 Grenoble, France
5 Institut de Minéralogie, de Physique des Matériaux et de Cosmochimie, Sorbonne Université — UMR CNRS 7590, F-75005 Paris, France
6 Synchrotron SOLEIL, L'Orme des Merisiers Saint-Aubin, BP 48 F-91192 Gif-sur-Yvette Cedex, France
7 Université de Lorraine, CNRS, IJL, F-57000, Metz, France
8 Université de Lorraine, CentraleSupélec, LMOPS, F-57000 Metz, France

([*] Author to whom correspondence should be addressed: olivier.pages@univ-lorraine.fr)

**Abstract**
Alloying offers a convenient route to tune the physical properties of semiconductors, at the price, however, of local lattice/bond distortions needed to accommodate the lattice mismatch. Being nearly lattice-matched and moreover highly resistant to lattice/bond distortions thanks to a largely covalent bonding, cubic-$Zn_{1-x}Mn_xTe$ sets a benchmark to test the sensitivity of lattice vibrations to minimum lattice/bond distortions induced by alloying. We explore experimentally the two transverse optical regimes of cubic-$Zn_{1-x}Mn_xTe$ ($x \leq 0.7$), with *ab initio* calculations in support: (i) the phonon-polariton regime at quasi-infinite (photon-like) wavelengths "averaging" over the alloy disorder, probed by forward-Raman scattering, and (ii) the purely-mechanical phonon regime down to minimum wavelengths nearly matching the lattice constant "zooming" into the alloy disorder, achieved by inelastic neutron scattering. The short Zn-Te bond is found to exhibit a compact optical-duo distinguishing *homo* from *hetero* local bond environments (1-bond→2-mode), as explained by the percolation model. This deviates from the historical $Zn_{1-x}Mn_xTe$ MREI (modified-random-element-isodisplacement) scheme inspired from "classical" Raman and infrared studies, in which the optical modes are "blind" to the bond environment (1-bond→1-mode). So, the test is positive, reinforcing the status of the percolation scheme as a robust decryptor of (even small) lattice distortions induced by alloying through the lattice dynamics.

The disordered (high-entropy) magnetic II-VI semiconductors alloys (SCA) involving the magnetic $Mn^{2+}$ ion in substitution, such as $Zn_{1-x}Mn_xTe$ with cubic/zincblende-zb structure ($x \leq 0.7$), are attractive for spintronics due to their capacity to combine, through alloying, magnetic and optical properties[1]. A major drawback, however, is that alloying causes local lattice distortions, needed to accommodate the contrasts in atom/bond physical properties (atomic mass, atomic size, bond length-$l$, bond force constant-$k$). The lattice distortions fragilize the crystal and degrade the physical properties, limiting applications[2].

The lattice dynamics offers a suitable playground to explore the lattice distortions at all length scales since the wavelength ($\lambda$) of lattice vibrations ranges from quasi infinite near the Brillouin zone center $\Gamma$ (BZC, $q = 2\pi \times \lambda^{-1} \rightarrow 0$), "averaging" over the alloy disorder, all the way down to minimum of twice the lattice constant ($a$) near the Brillouin zone boundary (BZB, $\lambda \sim 2 \times a$), "zooming" into the alloy disorder.

For probing the lattice distortion of SCAs through the lattice dynamics the optical modes seem preferable over the acoustic modes. The latter are the genuine collective lattice vibrations, dragging all bond species of a SCA into the same dynamics, *i.e.*, at the same frequency[3,4]. As the acoustic modes already fail to distinguish bond species, *a fortiori* they presumably fail also to distinguish local environments of a given bond, being hence "double-blind", *i.e.*, bond-insensitive *and* environment-insensitive. In contrast, the optical modes, inherently originating from effective bond-stretching, offer a local (bond-related) insight into the lattice distortions. One may moreover hope that the optical modes are not only bond-sensitive but also environment-sensitive. The latter virtue is denied by the historical modified-random-element-isodisplacement (MREI) model[5] set at the emergence of zb-SCAs in the 1960s. Accordingly, the like bonds of a given species vibrate at the same frequency at a given composition $x$ (1-bond→1-mode), irrespectively of the local environment. One may say, in a nutshell, that the MREI optical modes are "single-blind", *i.e.*, bond-sensitive *but* environment-insensitive. In our percolation model, designed in the 2000s[6], the optical modes "see everything", distinguishing *homo* from *hetero* environments *per* bond (1-bond→2-mode). The cluster model[7] promoted the same view ahead of time, distinguishing like bonds according to their first-neighbor environment, among four possible ones per bond in a zb-SCA (1-bond→4-mode). However, to our view the cluster model is biased[8,9] – hence not further discussed below – by a 1D/3D ambivalence that the equations of motion per atom are scalar (one-dimensional, 1D) while the elementary oscillators behind the individual Raman/IR features, *i.e.*, the first-neighbor clusters, are defined in the real (three-dimensional, 3D) crystal.

To elucidate the MREI vs. percolation nature of the lattice dynamics of a SCA, we focus on the transverse optical (TO) modes ($\vec{u} \perp \vec{q}$), due to atom vibrations ($\vec{u}$) perpendicular to the direction of propagation (given by the wavevector $\vec{q}$). The TO modes exhibit a phonon-polariton (PP) regime[10,11] resulting from phonon-photon coupling at quasi infinite (photon-like) wavelengths near $\Gamma$ ($\lambda \rightarrow 10^4 \times a$) plus a mechanical (mTO) phonon regime[10,11] ruling over the BZwide rest ($10^2 \times a > \lambda > a$) – methods. The TO variety contrasts with the basic invariance of the electrical/mechanical longitudinal optical (LO, $\vec{u} \parallel \vec{q}$) phonon[10,11] (methods), offering the possibility to explore the alloy disorder from different (PP + mTO) angles. The PP and mTO regimes are probed by forward inelastic visible-light/Raman scattering[11] (forward-RS, schematically operated in "transmission") using lasers and by inelastic neutron[12] (INS) or X-ray[12] (IXS) scattering using nuclear reactors and synchrotrons, respectively. The maximum $q$ achieved by RS ($\lambda \sim 10^2 \times a$), *i.e.*, in backward scattering (backward-RS, schematically operated in "reflection")[11], matches the minimum $q$ achieved by INS/IXS[12]. So, RS and INS/IXS are complementary and not competing.

The percolation model successfully applied to all tested SCA so far, with cubic (zincblende[6], diamond[13]) and hexagonal (wurtzite[14]) structures, suggesting its universality. The emphasis was placed on mismatched SCA with the idea in mind, verified in the case studies, that a large mismatch ($\frac{\Delta a}{a}$) would clarify the percolation TO-duo per bond. Here, we proceed the other way around and test the percolation model on zb-$Zn_{1-x}Mn_xTe$ that features an ultimately positive case of alloying. The Zn and Mn substituents stand close by on the same line of the periodic table and thus have comparable atomic masses ($\frac{\Delta m}{m}\sim$15%) and covalent radii ($\frac{\Delta R_C}{R_C}\sim$12%). Remarkably, the light Mn atom is the big one. That matters for the percolation scheme (cf. infra). Besides, the lattice mismatch hits minimum ($\frac{\Delta a}{a}\sim$2.75%, Mn-Te being the long species – cf. infra)[15] while the bond covalency hits maximum, falling between the II-VI and III-V values for Zn-Te[16,17], and matching the III-V values for Mn-Te[18]. Generally, the bond-stretching and bond-bending force constants (symbolized by $k$) increase with the bond covalency[19]. And along with it, the resistance to bond distortions in shear[16], inherent to alloying and so destabilizing for the lattice. In brief, being well-matched and highly covalent, zb-$Zn_{1-x}Mn_xTe$ is subject to minimum lattice/bond distortions and is moreover highly resistant to lattice/bond distortions. As such, it features a benchmark to trace the onset of the percolation TO-duo per bond in SCAs back to a slight difference in bond relaxation/distortion depending on the local *homo*/*hetero* environment[6].

The state of the art about the zb-$Zn_{1-x}Mn_xTe$ lattice relaxation/dynamics summarizes as follows.

Mn incorporation achieves up to ~70 at.% in the zb-lattice of ZnTe[1], differing from the NiAs-lattice of MnTe[20]. As apparent in extended X-ray absorption fine structure (EXAFS) data ($0.15\leq x \leq0.65$)[22], the parent-like Zn-Te and Mn-Te bond lengths ($l$) are preserved at any $x$ (within error bars), as in most zb-SCAs[21], and also the nominal tetrahedron angle (109.8°) of the regular zb-structure, even at maximum alloying ($x\sim0.5$). A bond softens ($k$ ↘) when it is stretched ($l$ ↗), and vice versa. Along this naive one-to-one "lattice relaxation vs. lattice dynamics" 1D-correspondence, the $l$-invariant 1-bond→1-$l$ EXAFS picture for zb-$Zn_{1-x}Mn_xTe$ transposes into a $k$-invariant 1-bond→1-$k$ MREI scheme for its mTO modes.

Earlier studies of the zb-$Zn_{1-x}Mn_xTe$ lattice dynamics (Fig. 1), listed below in chronological order, were conducted in-house using optical techniques, *i.e.*, backward-RS and infrared-reflectivity (IR). By force of the quasi vertical dispersion of the visible (RS) and infrared (IR) light sources, both techniques operate near the BZC, specifically addressing the mTO and LO modes at $q$ values of the order of one percent of the BZsize[11] ($\lambda\sim10^2\times a$). Nothing whatsoever is yet produced about the zb-$Zn_{1-x}Mn_xTe$ optical modes at lower (PP-like) or higher (BZB-like) $q$ values.

The pioneering IR[23,24] and RS[24-26] studies, done at $x\leq0.15$, pointed towards a 2-mode MREI pattern involving a large Zn-Te mTO-LO band beneath a narrow Mn-Te one, the two arranged into a compact mTO-LO-mTO-LO sequence. This is highly intricate due to a strong overlapping of the parent mTO-LO bands, the RS/IR frequencies matching within 4 $cm^{-1}$ (Ref. 27). The phonon intricacy is aggravated by the Mn-impurity mode in zb-ZnTe (which marks the closure of the Mn-Te TO-LO band at $x\sim0$) being quasi degenerated with the matrix-like Zn-Te LO mode (within 2 $cm^{-1}$). A similar phonon intricacy, complicating the phonon assignment, occurs with zb-$In_{1-x}Ga_xAs$[28] and zb-$Cd_{1-x}Zn_xTe$[29]. By increasing $x$ the Mn-Te band expands while the Zn-Te band shrinks, as clarified by RS thanks to a breakthrough in crystal growth opening the entire zb-domain ($x\leq0.7$)[27]. Notably, the lower LO mode suffers a major softening (~20 $cm^{-1}$). All alternative optical modes are hardly $x$-sensitive. Only, the m-TO branches slightly bend downward (by ~10 $cm^{-1}$) in a $\omega(x)$-plot (Fig. 1, $\omega$ is the phonon frequency), due to the macroscopic tensile strain suffered by both bond species when the lattice expands by increasing $x$. Generally, the reported mTO Raman modes in the literature are obscured by LO modes nearby, rendering it difficult to appreciate any percolation-type mTO fine structure.

In this work we test whether the minimum lattice/bond distortions suffered by zb-$Zn_{1-x}Mn_xTe$ suffice, or not, to generate any percolation TO-duo per bond beyond the admitted MREI description[27]. The test is done by specifically addressing experimentally the TO modes in their full (PP + mTO) diversity using high-quality zb-$Zn_{1-x}Mn_xTe$ ($x\leq0.70$)[27] single crystals grown by the Bridgman technique (methods, Fig. S1 – the prefix S stands for supplementary material). The mTO regime is studied not only in backward-RS near the BZC ($\lambda\sim10^2\times a$)[11] but also by INS near the BZB ($\lambda\sim2\times a$) in hope that a percolation mTO-duo for $Zn_{1-x}Mn_xTe$, if any, will be best resolved at the latter limit, as observed with cubic-$Zn_{0.67}Be_{0.33}Se$[30] and hexagonal-$Zn_{0.67}Mg_{0.33}S$[14]. Forward-RS is further used to probe the PP dispersion from the deep photon-like regime ultimately close to the BZC (forward-RS, $\lambda \to 10^4 \times a$) all the way back to the native mTO-regime of the PPs (backward-RS, $\lambda \sim 10^2 \times a$)[11]. The experimental data are discussed using our generic $q$-dependent PP-RCS set within the linear dielectric formalism (specifically using Eq. 1 of Ref. 29), covering the PPs together with their asymptotic mTO and LO modes near the BZC ($10^4 \times a \geq \lambda \geq 10^2 \times a$), with *ab initio* (SIESTA[31]) mTO-phonon calculations near the BZC ($\lambda \sim 10^2 \times a$) at $x \sim 0$ in support.

## Results and discussion

A basic rule is that the percolation TO-duo is best resolved for the small-$R_c$ substituent, hence typically the light one that vibrates at high frequency[6]. This is because the small substituent has more room to accommodate locally the lattice mismatch, with concomitant impact on its TO frequencies, being more diversified. $Zn_{1-x}Mn_xTe$ is a special case in that the small-$R_c$ substituent, *i.e.*, Zn, is the heavy one, vibrating a low frequency. This points towards a three-TO $\{2 \times (Zn - Te), 1 \times (Mn - Te)\}$ percolation scheme.

**Mechanical-TO (m-TO) phonon regime**

In principle, the native mTO regime of the PPs is best suited to elucidate the TO mode behavior of a SCA. This is because mechanical modes hardly couple and thus preserve the natural phonon diversity of a complex system like a SCA. The mTO modes of pristine zb-ZnTe[32,33] and zb-MnTe[34] hardly disperse across the BZ (by few cm$^{-1}$ at most), suggesting the same for zb-$Zn_{1-x}Mn_xTe$. The Anderson's criterion for phonon localization[35] is thus easily fulfilled. Hence, even minor changes in bond distortion, *e.g.*, those depending on the local *homo*/*hetero* bond environment in the spirit of the percolation model, are likely to generate distinct mTO modes per bond.

Pure-mTO $\Gamma$-like ab initio calculations

A pure-mTO *ab initio* insight near $\Gamma$ at $x \sim 0$ suffices to fix ideas. An isolated Mn atom in a large (64-atom) zb-ZnTe supercell ($x \sim 0$) generates a local compression because Mn-Te is longer than Zn-Te. The Zn-Te bonds near Mn are thus shorter than the matrix-like Zn-Te bonds (Fig. 2b) and hence create a minor/local m$TO_{Zn-Te}^{Mn}$ mode due to Zn-Te vibrations (subscript) in the *hetero* Mn-environment (superscript) at a higher frequency than the dominant/matrix m$TO_{Zn-Te}^{Zn}$ mode due to Zn-Te bonds away from Mn vibrating in *homo* environment (Fig. 2a, inset). Small as it is (~5 cm$^{-1}$), the m$TO_{Zn-Te}^{Mn}$-m$TO_{Zn-Te}^{Zn}$ frequency gap is a priori visible by RS (~0.3 cm$^{-1}$) and possibly also by INS away from $\Gamma$[14,30] (~13 cm$^{-1}$ at ~160 cm$^{-1}$~20 meV), thanks to the high spectral resolutions of both techniques, as specified in brackets.

Backward Raman scattering

Pure-mTO zb-$Zn_{1-x}Mn_xTe$ backward-RS spectra display separate Zn-Te (lower) and Mn-Te (upper) MREI-like signals (Fig. 2c). The Zn-Te duo and the Mn-Te mode are genuine mTO modes at any $x$, as was verified by symmetry analysis in crossed/parallel polarizations (Figs. 2d-S2), further attesting the high vibrational quality of our crystal series. The symmetry analysis is not so positive at maximum alloying (not shown). So, $x$=0.51, while punctually used hereafter, is not emphasized below.

Looking closely, the Zn-Te signal around maximum zb-alloying ($x$=0.51, 0.67) seems bimodal. According to the *ab initio* insight (Fig. 2a), the lower and upper modes are assigned as m$TO_{Zn-Te}^{Zn}$ and m$TO_{Zn-Te}^{Mn}$, respectively. The Raman intensities scale as ~2:1 at $x$=0.51 and ~1:1 at $x$=0.67, indicating a sensitivity to *homo* (pure-Zn + Zn/Mn-mixed) and *hetero* (pure-Mn) environments up to second-neighbors[9,29], assuming a random Zn/Mn substitution. As $x$ gets smaller, the related oscillator fractions ($f$-terms) massively turn to the advantage of the lower-m$TO_{Zn-Te}^{Zn}$ mode. The Zn-Te duo nevertheless persists in the form of a slightly asymmetrical Zn-Te lineshape at $x$~0.3 (the intensity ratio is ~6:1) but disappears at $x$~0.1 (~50:1). This is consistent with the only reported pure-mTO $Zn_{1-x}Mn_xTe$ RS spectrum in the literature, taken by Peterson *et al*.[27] at $x$=0.19 (Fig. 2b therein), showing no trace of the Zn-Te duo.

In brief, there are positive signs of a Zn-Te mTO duo near the BZC ($\lambda \sim 10^2 \times a$) at moderate/large Mn content. However, the duo is so compact that it can hardly emerges in *ab initio* nor in experiment (backward-RS), unless one attentively concentrates on it.

Inelastic neutron scattering

The situation might improve by INS. Indeed, the percolation Zn-Te mTO-duo being due to a slight difference in local lattice relaxation around Zn-Te depending on its *homo/hetero* environment, it should be best resolved by INS at minimum $\lambda$ ($\sim 2 \times a$) "zooming" into the local alloy disorder than by RS at large $\lambda$ ($10^4 \times a > \lambda > 10^2 \times a$) "averaging" over the alloy disorder[14,30]. The Zn-Te mTO-duo, while compact when probed by RS near the BZC ($\lambda \sim 10^2 \times a$, Fig. 2e, bottom curve), broadens in INS data acquired along $[00\bar{1}]$ at $\sim$10% of the BZsize ($\lambda \sim 10 \times a$, central curve) and eventually divides into a distinct duo near the BZB ($\lambda \rightarrow a$, top curve), as expected. Small error bars due to large neutron-counting times (six hours per spectrum) ascertain the trends, clarified by two- ($\sim$BZC) and three-Gaussian (BZB) fits (solid curves) of the neutron data (symbols). Nevertheless, although the INS data point toward a change in paradigm from the MREI model to a percolation picture for $Zn_{1-x}Mn_xTe$, they should be regarded as indicative only, since the relatively large sample ($\sim$75 $mm^3$) investigated by INS does not exhibit the same high structural quality as the smaller crystal pieces (a few $mm^3$) used for RS (methods).

**Phonon-polariton (PP) regime**

Additional experimental insight is searched for by departing from the mTO-phonon regime probed by INS/backward-RS and exploring the PP regime by forward-RS. Advantage is taken that the small-$R_c$ substituent (Zn) supporting the best-resolved TO-duo in principle (cf. supra) vibrates at low frequency. In this rare case (given that the small-$R_c$ substituent is generally the light one vibrating at high frequency), the TO-duo, however compact it may be in the mTO-phonon regime away from $\Gamma$, breaks away on entering the PP regime towards $\Gamma$. Basically, the lower submode dramatically softens by converging onto the quasi vertical photon asymptote, whereas the upper submode remains pinned within the native m-TO band. In Fig. 3, an overview of the zb-$Zn_{0.63}Mn_{0.37}Te$ $q$-dependent PP-RCS involving a percolation mTO-duo reflecting a sensitivity of Zn-Te vibrations to second neighbors, is illustrative with this respect. Both the Raman frequencies (curves) and the Raman intensities (thickness of curves) are indicated.

To probe the kind of PP-RCS shown in Fig. 3 nothing yet can be expected from *ab initio* calculations due to the difficulty in handling the macroscopic $\vec{q}$-dependent photon-like PP electric field. Prospective *ab initio* calculations of the PP dispersion (frequency aspect) are underway for pristine cubic-GaP and hexagonal-GaN[36], however, not addressing the Raman efficiency (intensity aspect). INS/IXS fall also short of addressing PPs due to a common selection rule limiting the INS/IXS insight to the sole longitudinal modes within the first BZ, where the photon dispersion develops[12] and hence also the PP coupling. The point is that PPs are transverse modes[11]. Forward-RS is the only option left, yet unused to study the PPs of SCAs, apart in our own recent works (Ref. 29 and refs. therein).

At any $x$, cross-polarized zb-$Zn_{1-x}Mn_xTe$ forward-RS spectra taken with the 632.8 nm laser line manifest three novel features (Figs. S3-S4), absent in backward-RS. The three consist of genuine PPs since they disperse and strengthen/collapse by changing $\theta$. The TO character is further attested by the three modes obeying the same forward-RS selection rules (Figs. 4-S4) as their native Zn-Te and Mn-Te mTO modes probed in backward-RS (Fig. S2). The lower-$PP^{-}$ and upper-$PP^{int+}$ modes refer to the heavy Zn-Te and light Mn-Te bonds, respectively. The intermediary-$PP^{int-}$ mode disperses near the Zn-Te mTO band, hence ZnTe-like. Altogether, the three PPs support an effective change of paradigm from MREI/2-mode to percolation/3-mode for zb-$Zn_{1-x}Mn_xTe$ ($x \leq$0.7).

PP-coupling

However, the observed PPs exhibit odd attributes. First (anomaly 1), the high-frequency (superscript) $PP^{int+}$ (probed at $\theta_{min}$~0°) and $LO^{+}$ ($\theta$-independent) modes are sharper than their native Zn-Te and Mn-Te mTO modes (probed at $\theta_{max}$~180°). Second (anomaly 2), at $\theta_{min}$ $PP^{int-}$ and $PP^{int+}$ emerge at lower frequencies than predicted by theory (PP-RCS) even in perfect forward-RS ($\theta$=0°, Fig. 4 – double-arrows). This happens to such extent that $PP^{int-}$ even disperses (slightly) outside the Zn-Te mTO duo/band, and not inside it (anomaly 3). It is just the opposite for $LO^{+}$ (a variant of anomaly 2), showing up at (slightly) higher frequency in experiment than predicted by theory (LO-RCS, Fig. S3).

Anomaly 1 (linewidth) cannot be explained by a breaking of the translational symmetry due to the alloy disorder, leading to the contribution of $q \neq 0$ modes to the Raman signal, in the spirit of the spatial correlation model[37], as done, *e.g.*, for $Al_{1-x}Ga_xAs$, another well-matched zb-SCA[3]. Indeed, the mTO and LO modes of zb-ZnTe are hardly dispersive and much dispersive, respectively[32,33]. So, involving $q \neq 0$ modes besides the nominal Raman active $q$=0 modes would leave the mTO mode sharp and broaden the LO mode. This is just the opposite for zb-$Zn_{1-x}Mn_xTe$. In fact, it was argued that the spatial correlation model may not apply in case of alloying, the Raman signal primarily referring to $q$=0[38].

The conjunction of anomalies 1 (linewidth) and 2 (frequency) affecting $PP^{int+}$ and $LO^{+}$ cannot be merely fortuitous, suggesting a common origin. Notably, anomaly 1 marks a difference between, on the one hand, $PP^{int+}$ and $LO^{+}$, carrying electric fields ($E_{T,L}$, the subscript standing for transverse-PP and longitudinal-LO), and, on the other hand, their native purely-mechanical mTO modes. Hence, anomaly 1 relates to $E_{T,L}$. To appreciate how anomaly 1 develops with $q$, $PP^{+}$ is a better candidate than $LO^{+}$ since $E_T$ carried by $PP^{int+}$ is $q$-dependent whereas $E_L$ accompanying $LO^{+}$ is permanent. Basically, $E_T$ achieves maximum in the deep PP regime probed by forward-RS near $\Gamma$ at $\theta \rightarrow 0$ and vanishes to zero by deviating a few degrees from the strict normal incidence/detection, which suffices to off-shift the TO modes away from $\Gamma$ towards the native mTO regime of the PPs.

Anomaly 1 affecting the Mn-related $PP^{int+}$ mode is especially clear at large Mn content ($x$=0.37, 0.67 – Figs. 4-S3). A close examination of the $\theta$-dependent forward-RS spectra reveals that $PP^{int+}$ suffers an abrupt, and not progressive, broad→sharp transition on crossing the photon-like $PP^{int+}$ extinction near the S-inflexion, at which limit $E_T$ suddenly takes on a macroscopic (photon-like) character. Basically, $PP^{int+}$ is broad *ante* extinction (away from $\Gamma$), as its native mTO mode, and sharp *post* extinction (near $\Gamma$). Notably, just beyond the S-like extinction, $PP^{int+}$ is fast (photon-like), well-supported by the lattice (phonon-like, attested by a significant Raman efficiency) and has a long lifetime (sharp). This is interesting for photonics in view to transfer information at high velocity through the lattice in the THz.

To our view, anomaly 1 reflects an intramode transfer of oscillator strength mediated by $E_T$. This presupposes some fine structuring of the native mTO modes of the $PP$s. The fine structure originates from local $x$-fluctuations inherent to alloying, impacting the Raman frequencies. *Ante* extinction, $E_T$ has not yet developed into a macroscopic field and remains local. So, the elementary $PP^{int+}$ modes due to the individual mTO submodes forming the fine structure hardly $E_T$-couple. In this "uncoupled" case, the overall $PP^{int+}$ ($E_T \neq 0$) and mTO ($E_T$=0) modes exhibit similar fine structures "echoing" each other. Hence, the $PP^{int+}$ and mTO modes are equally broad. *Post* extinction, the elementary $PP^{int+}$ submodes couple via $E_T$ that has turned macroscopic. The net effect of the $E_T$-coupling is to channel most of the available oscillator strength towards the lower $PP^{int+}$ oscillator of the series, becoming giant and downshifted beneath the series[39]. It is just the opposite for the $E_L$-coupling, the upper oscillator being favored and upshifted[39,40]. As $PP^{int+}$ at $\theta_{min}$ and $LO^{+}$ refer to a unique mode of the series, they are both sharp, explaining anomaly 1. Anomaly 2 affecting $PP^{int+}$ and $LO^{+}$ is due to their respective downward and upward shifts.

The reported $Zn_{0.63}Mn_{0.37}Te$ $PP^{int+}$ simulations for the 632.8 nm laser line in Fig. 3 (top curves) illustrate how the $E_T$-coupling develops by reducing $\theta$. A discrete Mn-Te mTO fine structure is involved, arbitrarily comprising three equally-spaced sub-mTO oscillators – sufficient to fix ideas – sharing the available Mn-Te oscillator strength in proportion to 30%-40%-30% (right, black curve). *Ante* extinction ($\theta \nearrow$) $PP^{int+}$ divides into three distinct submodes with similar intensities in echo to the native three distinct $mTO_{Mn-Te}$ submodes. *Post* extinction, the lower $PP^{int+}$ submode attracts most of the available Mn-Te oscillator strength. At $\theta_{min}$ this $PP^{int+}$ submode (top, black curve) emerges below (double-arrows) the $PP^{int+}$ mode generated in absence of $mTO_{Mn-Te}$ fine structuring (red curve).

At $x$~0.37, any $mTO_{Mn-Te}$ "manipulation" (absence/presence of fine structure) leaves the $LO^+$ frequency unchanged (Figs. 3-4). However, anomaly 2 (upward shift) affecting the sharp $LO^+$ mode becomes appreciable at $x$~0.67 corresponding to similar Zn-Te and Mn-Te mTO Raman intensities (Fig. S3). This supports an effective fine structuring of the latter two (broad) features from the LO point of view, in line with the PP point of view.

One may well wonder why $PP^-$ related to Zn-Te is not as sharp as $PP^{int+}$ related to Mn-Te, since the native mTO modes of $PP^{int+}$ and $PP^-$, *i.e.*, $mTO_{Mn-Te}$ and $mTO_{Zn-Te}^{Zn}$, respectively, are equally broad, hence subject to similar multi-mode decompositions. This $PP^-$ vs. $PP^{int+}$ discrepancy is due to an experimental limitation of our forward-RS setup that the laser-focusing ($\vec{k}_i$) and light-collecting ($\vec{k}_s$) cones have finite angles. This generates a continuous distribution of scattering angles $\theta = (\widehat{\vec{k}_i, \vec{k}_s})$ inside the crystal around the mean $\overline{\theta}$ defined by the mean $\vec{k}_i$ and $\vec{k}_s$ along the cone axis. A typical $\theta$-uncertainty of $\Delta\theta \sim \pm 0.5$ has been estimated for our forward-RS setup with hexagonal-GaN[41]. In turn $\Delta\theta$ creates a $\Delta\omega$ uncertainty on the $PP$ frequency. Its magnitude depends on the slope of the probed portion of the $\omega$ vs. $q$ $PP$ dispersion at $\overline{\theta}$. $PP^{int+}$ probed at $\theta_{min}$ is sharp ($\Delta\omega \rightarrow 0$) because $PP^{int+}$ is quasi dispersionless *post* the photon-like extinction where $\theta_{min}$ operates (Fig. 3). In contrast, $\theta_{min}$ probes the bottleneck of the $PP^-$ dispersion on entry to the deep photon-like regime where the $PP^-$ dispersion is so steep (Fig. 3). So, even if unimodal (cf. supra), the $PP^-$ mode at $\theta_{min}$~0°, integrated over $\Delta\theta$, happens to be broad.

PP-decoupling

Anomaly 3 concerned with $PP^{int-}$ is a separate case. By no way a simulation based on a fine structuring of the native mTO modes, so successful to explain anomalies 1 and 2 affecting $PP^{int+}$ and $LO^+$, can generate a $PP^{int-}$ dispersion outside the Zn-Te mTO band (Fig. 4b, double-arrows). Various attempts in this line (not shown), playing with the number and spacing of oscillators forming the mTO fine structure, repeatedly turned out to be a sterile exercise.

To our view the slight *homo*/*hetero* difference in local lattice distortions around the Zn-Te bonds, at the origin of the $mTO_{Zn-Te}^{Zn} - mTO_{Zn-Te}^{Mn}$ percolation mTO-duo, suffices to prevent an effective $E_T$-coupling between the related $PP^-$ and $PP^{int-}$ modes. Hence, albeit the $mTO_{Zn-Te}^{Zn}$ and $mTO_{Zn-Te}^{Mn}$ modes are quasi degenerate (Fig. 2), they engage the PP regime on their own as uncoupled modes. Retrospectively, the PP-decoupling is a positive surprise. By preserving their natural oscillator strengths, without $E_T$-mediated pooling, $PP^{int-}$ and $PP^-$ remain visible at any $\theta$. Strictly speaking, due to their uncoupling, $PP^{int-}$ and $PP^-$ should rather be labeled in terms of $PP_{Zn-Te}^{Mn}$ and $PP_{Zn-Te}^{Zn}$ in accordance with their direct $mTO_{Zn-Te}^{Mn}$ and $mTO_{Zn-Te}^{Zn}$ ascendants, respectively. However, the original notation is maintained since the latter two ZnTe-like modes, if not coupled to each other, do separately couple with the MnTe-like mode on entering the PP-regime.

At moderate Mn content ($x$=0.12, 0.37), $mTO_{Zn-Te}^{Zn}$ is awarded most of the available Zn-Te oscillator strength, and $mTO_{Zn-Te}^{Mn}$ gets very few of it (Fig. 5a). So, the related $PP^-$ and $PP^{int-}$ modes at $\theta_{min}$ disperse far below the native mTO band and very close to it, respectively (Fig. 4a). At large Mn content ($x$=0.67), the Zn-Te oscillator strength shares nearly equally between both native mTO submodes. In this case the $PP^-$ and $PP^{int-}$ downshifts at $\theta_{min}$ are comparable (Fig. S3).

The $Zn_{1-x}Mn_xTe$ PP-RCS at $\theta_{min}$=0° calculated for the 632.8 nm laser line by assuming a Zn-Te sensitivity to *homo*/*hetero* environments to second-neighbors (methods) fairly support the PP-uncoupling at $x$=0.37 and $x$=0.67 (Fig. 4).

**LO phonon regime**

We cannot escape a brief discussion of the $E_L$-coupling. This generates massive LO distortions apparent in Fig. 5 displaying tilted RCS views (three-dimensional, best suited to appreciate the Raman intensities in their $x$-dependence) of the $\{mTO_{Zn-Te}^{Zn}, mTO_{Zn-Te}^{Mn}, mTO_{Mn-Te}\}$ (a), uncoupled-$\{LO_{Zn-Te}^{Mn}, LO_{Zn-Te}^{Zn}, LO_{Mn-Te}\}$ (b) and coupled-$\{LO^-, LO^{int}, LO^+\}$ (c) modes of zb-$Zn_{1-x}Mn_xTe$ virtually expanded over 0≤ $x$ ≤1. Frontal views (two-dimensional, more appropriate to compare the Raman frequencies in their $x$-dependence) are provided in supplementary material (Fig. S5, top). The same set of input parameters, notably a small phonon damping (1 $cm^{-1}$) to distinguish neighboring modes, is used throughout. The same notation with a subscript (vibrating bond) and a superscript (*homo*/*hetero* environment) is used to label the native naturally-uncoupled mTO modes and their artificially-uncoupled LO-descendants.

Dramatic $E_L$-mediated $LO_{Zn-Te} \rightarrow LO_{Mn-Te}$ inter-bond ($x$~0) and intra-bond/duo (Zn-Te) $LO_{Zn-Te}^{Zn} \rightarrow LO_{Zn-Te}^{Mn}$ ($x$~0.5) transfers of oscillator strength (intensity aspect) manifested by strong repulsions (frequency aspect) occur at the resonances corresponding to frequency matching of the related uncoupled-LO modes, as specified in brackets. The net effect of the $E_L$-coupling is to create a giant $LO^+$ mode with quasi stable intensity and frequency in the Mn-Te spectral range above a minor $LO^{int}$ mode that collapses downward the Zn-Te spectral range by increasing $x$. The remaining $LO^-$ oscillator, quasi deprived of oscillator strength, hence not visible experimentally, repelled close to the lower $mTO_{Zn-Te}^{Zn}$ mode, plays a minor role.

So, in first approximation a MREI model of two $\{LO_{Zn-Te}, LO_{Mn-Te}\}$ $E_L$-coupled harmonic oscillators (Fig. S5, bottom) suffices to describe the inter-bond LO coupling, as proposed for zb-$Cd_{1-x}Zn_xTe$ and zb-$Ga_{1-x}In_xAs$[29]. This comes to merge the $mTO_{Mn-Te}$–$mTO_{Zn-Te}^{Zn}$ percolation mTO-duo into a $mTO_{Zn-Te}$ MREI-singlet with oscillator fraction $f_{Zn-Te}(x) = 1 - x$. Comparative insights into the MREI (2-mode, black/right-oriented) and percolation (3-mode, red/left-oriented) mTO (thick curves) and LO (thin curves) RCS at the experimentally studied $x$ values are shown in Fig. 1. By diagonalizing the dynamical matrix and expressing the unit perpendicular eigenvectors from $|LO_+\rangle = \binom{\cos\delta}{\sin\delta}$, $\cos^2\delta$ and $\sin^2\delta$ estimate how much in percent the Zn-Te and Mn-Te bonds are involved in the upper ($LO^+$) coupled mode depending on $x$, and vice versa for the lower ($LO^-$) one. $\tau=\tan\theta$ is a direct marker of bond-mixing. Not surprisingly, the LO modes are (quasi) decoupled (bond-specific, $\tau \rightarrow 0$) at $x \rightarrow 1$, *i.e.*, away from the resonance, and strongly coupled (bond-mixed, $\tau \rightarrow \tau_{max}$) at $x \rightarrow 0$, *i.e.*, near the resonance (Fig. S5). Remarkably, zb-$Zn_{1-x}Mn_xTe$ achieves a higher bond-mixing ($\tau_{max}$~0.5 at $x$~0.10) than zb-$Cd_{1-x}Zn_xTe$ and zb-$Ga_{1-x}In_xAs$ ($\tau_{max}$~0.4 at $x$~0.05)[29].

## Conclusion and perspectives

Cubic-$Zn_{1-x}Mn_xTe$ ($x$ ≤0.7) offers a benchmark to test the sensitivity of the TO-percolation model[6] to minimum lattice/bond distortions induced by alloying. A Zn-Te TO-duo marking a difference in lattice relaxation around Zn-Te depending on its local *homo*/*hetero* environment, in the spirit of the percolation model, is evidenced in the mTO regime. However, the mTO-duo is so compact and the mTO regime so dispersionless across the BZ that the duo can hardly be resolved *ab initio* or experimentally, whether probing finite-small $q$ in backward-RS or maximum $q$ by INS. The situation clarifies on entry to the highly-dispersive PP regime at $q\rightarrow 0$, probed by forward-RS. The Zn-Te duo splits off, supporting an effective MREI[27]-$\{1 \times (Zn-Te), 1 \times (Mn-Te)\} \rightarrow$ percolation[6]-$\{2 \times (Zn-Te), 1 \times (Mn-Te)\}$ shift of paradigm for the optical modes of zb-$Zn_{1-x}Mn_xTe$. Odd attributes (linewidth, frequency, intensity) of the individual PP modes depending on $q$ are discussed in terms of the microscopic⟷macroscopic transition affecting their accompanying transverse (photon-like) electric field on crossing the photon resonance.

Along with disordered (high-entropy) semiconductor alloys (SCAs), the high-entropy (metallic) alloys (HEAs) are exemplary systems to study how the lattice dynamics is impacted by disorder. Their lattice dynamics, governed by the bond force constants, *i.e.*, a local (microscopic) mechanical property, attracts a special interest due to HEAs' outstanding (macroscopic) mechanical properties[42]. The acoustic modes of HEAs were extensively studied over the past few years[43-45]. However, the optical modes, despite their unique ability to probe the local alloy disorder (percolation scheme), did not. Either because the optical modes are just absent[43-45], or, on the contrary, because they present a too complex or confusing pattern[46] (depending on the number of atoms per primitive cell). Anyway, an insight into the capacities of the percolation scheme set from SCAs in relation to HEAs might be intriguing.

# Methods

## Samples

$Zn_{1-x}Mn_xTe$ single crystals at well-separated compositions $x$ ($x$=0, 0.12, 0.37, 0.51, 0.67) are grown from the melt by mixing high-purity ZnTe (5N quality) with Mn (2N quality) using the Bridgman method. The zb-structure is attested throughout by powder X-ray diffraction either done at ambient pressure in laboratory conditions using the 1.541 Å Cu K$\alpha$ line and/or at nearly ambient pressure at the PSICHÉ beamline of synchrotron SOLEIL using the 0.3738 Å radiation (Fig. S1). The lattice parameter $a$ depends linearly on $x$ (Fig. S1) measured (with accuracy better than 1 at.%) by inductively coupled plasma-optical emission spectroscopy[47], as expected in case of a regular Zn↔Mn substitution on the cationic site[48].

## *Ab initio* calculations

An *ab initio* insight into the capacity of the optical modes due to the Zn-Te bonds of zb-$Zn_{1-x}Mn_xTe$ to distinguish *homo* from *hetero* environments in the spirit of the percolation model, is achieved by applying the SIESTA code[31] to a large 2×2×2 (64-atom) ZnTe-like zb-supercell containing an isolated Mn-impurity ($x\sim0$, Fig. 2). The finite-displacements force constants calculation is done after full relaxation (lattice constant and atom positions). As no disposition was applied to long-range electric fields, the reported *ab initio* data are thus of the pure-mTO type, more precisely concerned with the mTO phonon density of states projected at $\Gamma$ ($\Gamma$-projected mTO-PhDOS), that assimilates with the pure-mTO Raman signal in first approximation.

## Inelastic neutron scattering (INS)

The mTO regime of zb-$Zn_{0.33}Mn_{0.67}Te$ is studied by INS at the thermal-neutron triple-axis IN22 of Institut Laüe Langevin at room temperature along the $[00\bar{1}]$ direction of the reciprocal space using a large crystal (~75 mm$^3$) with homogeneous composition (the Raman shifts across the sample surface are less than 1 cm$^{-1}$) aligned in the ($[00\bar{1}]$, [110]) horizontal scattering plane. IN22 was set up in a W configuration and inelastic scans were performed with constant $k_f$=2.662Å$^{-1}$, where $k_f$ is the wavevector of the scattered neutron beam. A pyrolytic graphite (PG(002)) monochromator and analyzer were used. Higher-order contamination was removed using a PG filter in the scattered neutron beam.

The mTO signal is specifically probed at ~10% of the BZsize, near the intense Bragg peak (331), *i.e.*, at (330.9), and exactly at the BZB, *i.e.*, (330). The neutron data (Fig. 2e, symbols with error bars corresponding to the square root of counts) were fitted with two (330.9) and three (330) Gaussian functions (after subtraction of oblique baselines), to appreciate the relative weights of the individual mTO-phonon modes. The linewidth per Gaussian function at a given transferred energy to the crystal ($E_i - E_f$) at least matches the spectral resolution $\delta$ at the magnitude $k_i$ of the corresponding wavevector of the incident neutron beam. For IN22, $\delta$ changes quasi linearly (with a slight negative bowing) from 1 meV to ~2 meV for ($E_i - E_f$) varying from 0 to 30 meV.

A meticulous in-plane (horizontal) scan at the (331) Bragg diffraction operated by (vertical) rotation of the crystal reveals a spurious Bragg diffraction due to a second minor crystal tilted by ~7° from the main one. Moreover, it is common that large $Zn_{1-x}Mn_xTe$ crystals grown by the Bridgman method are subject to twining[1], due to stacking faults in the ...*ABCABC*... sequence of hexagonal cation and anion close-packed planes along [111]. Hence, probing the main crystal at a selected ($hkl$) point of the (horizontal) scattering plane alternatively probes simultaneously ($h'k'l'$) points of the secondary and/or twined crystals, rendering the phonon assignment a difficult task. Even if an obstacle for a reliable study of the acoustic modes, this is not so critical for the currently studied mTO modes that are nearly dispersionless.

**Raman scattering (RS)**

Backward/forward-RS is done at ambient conditions using the 632.8 nm laser excitation at near-normal incidence/detection onto/through tiny (few $mm^3$) (110)-oriented crystal faces obtained by cleavage, corresponding to a pure-TO (TO allowed, LO forbidden) symmetry[49]. Generally, crossed polarizations of the incident laser ($\vec{e}_i$) and of the scattered light ($\vec{e}_s$) are used, to reduce the second-order signal obscuring the first-order TO (mTO, PP) one (Fig. S2). The TO symmetry is analyzed by rotating in pair the crossed (or parallel) polarizations at a given sample spot until completing a full revolution (360°) of the azimuth polarization angle ($\alpha$, step-increased by 20°) between $\vec{e}_i$ and the $[\bar{1}10]$ crystal axis taken as an arbitrary reference. The nominal TO (mTO or PP) Raman intensity (area of peak – modeled using a Lorentzian/Gaussian function – normalized to the area of the most intense peak of the series) vs. $\alpha$ variation describe Greekcross- and butterfly-like motifs for crossed ($\perp$) and parallel ($\parallel$)polarizations, respectively. The laser stability is checked by repeating the initial measurement at the end of the run.

The backward-RS and forward-RS geometries probe specific variants of the TO modes [11]. Due to the ionicity of the chemical bonding in a zb-SCA, its TO modes are likely to carry a transverse electric field $\vec{E}_T$. However, the latter can propagate only if $q$ for the TO mode (nearly) matches that for a photon, the pure transverse electric field, corresponding to $\lambda$ covering as much as $10^4$ crystal unit cells. Such small $q$ is achievable by forward-RS, minimizing the $\vec{q}$-transfer to the crystal. The probed TO mode in forward-RS is thus a phonon-photon coupled mode with mixed "electrical + mechanical" character, called a phonon-polariton (PP).

A single-oscillator pristine zb-crystal exhibits two $PP^-$ and $PP^+$ branches in repulsion with phonon and photon asymptotes. The latter describe the dispersion of light/photon in vacuum weighted by the static and high-frequency dielectric constants of the crystals on both sides on the phonon resonance[11], labelled $Ph^+$ and $Ph^-$ (Fig. 3), respectively. Alloying introduces multiple intermediary $PP^{int}$ branches (dividing from $PP^{int-}$ to $PP^{int+}$ from low to high frequency in case of multiple oscillators) with a characteristic S-like dispersion governed by two phonon asymptotes, mTO-like away from $\Gamma$ and LO-like near $\Gamma$. Generally, the PP modes suffer a Raman extinction on crossing ($PP^{int}$) or converging ($PP^-$ and $PP^+$) onto the photon-like asymptotes. This conforms to intuition that only matter-like excitations scatter light efficiently, whereas the photon-like ones do not.

Backward-RS achieves a maximum $q$-transfer to the crystal, *i.e.*, $q \sim 2 \times k_i$, falling far away from the $q(\omega)$ dispersion of a photon in the far-IR spectral range of a phonon. At this limit $\vec{E}_T$ cannot propagate. Hence, a TO mode probed in backward-RS reduces to a purely-mechanical phonon (mTO), deprived of electric field. It emerges that $\vec{E}_T$ is $q$-dependent. In contrast, the electric field $\vec{E}_L$ of a LO mode, being longitudinal, does not suffer any photon-like restriction. A LO mode thus keeps a uniform/permanent "electrical + mechanical" character at any $q$. Strictly at $\Gamma$, $\vec{E}_T = \vec{E}_L$ because at $q = 0$ there is no means to distinguish vibrations polarized perpendicularly to $\vec{q}$ (transverse) from vibrations polarized along $\vec{q}$ (longitudinal).

The same generic multi-oscillator (SCA-like) RCS, given by Eq. (1) of Ref. 29, derived within the linear dielectric formalism of Hon and Faust[50], is used to model the PP in their $q$-dependence ($q \rightarrow 0, \lambda \sim 10^4 \times a$) as well as their native mTO modes ($q \rightarrow \infty, \lambda \sim 10^2 \times a$) away from $\Gamma$ (being clear that RS operates near $\Gamma$ – main text) and their coupled- and uncoupled-LO modes near $\Gamma$ ($q \rightarrow 0, \lambda \rightarrow \infty$). The individual RCS due to a given uncoupled-LO/PP mode is obtained by removing all alternative oscillators from the LO/PP-RCS of zb-$Zn_{1-x}Mn_xTe$. AS for the regular multi-mode LO/PP-RCS, it naturally provides the coupled-LO/PP RCS.

Most input parameters of the RCS relate to the parent zb-ZnTe and zb-MnTe compounds. Together with $\varepsilon_\infty$, the mTO ($\omega_{TO}$) and LO ($\omega_{LO}$) Raman frequencies define the parent oscillator strengths $S = \varepsilon_\infty \times \omega_{TO}^{-2} \times (\omega_{LO}^2 - \omega_{TO}^2)$[5]. The used ($\varepsilon_\infty$, $\omega_{TO}$, $\omega_{LO}$, $C_{FH}$) values for zb-ZnTe and zb-MnTe are (7.3 – Ref. 24, 177.5 $cm^{-1}$ – Ref. 24, 207.0 $cm^{-1}$ – Ref. 24, -0.11 – Ref. 29) and (7.0 – Ref. 24, 188.0 cm$^{-1}$, 214.5 cm$^{-1}$, -6.65$\times 10^{-3}$), respectively, where $C_{FH}$ denotes the Faust-Henry coefficient. The last three values were adjusted in this work so as to best fit the experimental data. $C_{FH}$ for MnTe, not available in the literature, was adjusted so as to fairly replicate the experimental intensity ratio between the well-resolved Zn-Te and Mn-Te mTO backward-RS signals showing up with comparable intensities at $x$=0.67 (not shown).

The only alloy-related input parameters are $\omega_{T,p}(x)$ (apparent in Fig. 5a, the MREI description in the Zn-Te spectral range is linearly interpolated between the parent and impurity values – dotted curve) and $\gamma_p(x)$ of the mTOs directly observed in backward-RS experiments. Besides, $S_p(x)$ and $C_p(x)$ scale linearly with the fraction $f_p(x)$ of oscillator-$p$ in the crystal[51]. In the percolation-type three-oscillator $\{2 \times (Zn-Te), 1 \times (Mn-Te)\}$ description of zb-$Zn_{1-x}Mn_xTe$, the Zn-Te duo distinguishes purely-*hetero* Mn-Te environments at high-frequency from *homo*- and *homo*/*hetero*-mixed environments regrouped at low-frequency, as defined at the second-neighbor scale. The corresponding $f_p(x)$ are $(1-x)^3 + 2 \times x \times (1-x)^2$ and $x^2 \times (1-x)$, respectively[29]. For a MREI-like oscillator, that is blind to the local environment, $f_p(x)$ matches the corresponding bond fraction[29], *i.e.*, $x$ (Mn-Te) or $1-x$ (Zn-Te).

The $Zn_{1-x}Mn_xTe$ relative dielectric function $\varepsilon_r(\omega,x)$, governing the resonance term $\Delta(\omega,q,x) = \varepsilon_r(\omega,x) - q^2c^2\omega^{-2}$ of the RCS, plays a central role. A classical $\varepsilon_r(\omega,x)$ form is used comprising a linearly $x$-dependent electronic background $\varepsilon_\infty(x)$ besides Lorentzian $L_p(\omega,x)$ phonon mode contributions, *i.e.*, $p$ in total, depending on the used MREI- ($p$=2) or percolation-type ($p$=3) descriptions. Each $p$ mode is characterized by its mTO frequency, noted $\omega_{T,p}(x)$, its (sample-dependent) damping $\gamma_p(x)$ fixing the full width at half maximum of the Raman signal, and its oscillator strength $S_p(x)$ controlling both the magnitude of the $p$-related mTO – LO band and its absolute mTO Raman intensity. The RCS also involves the $p$-related Faust-Henry coefficients, noted $C_p(x)$, measuring the relative Raman scattering efficiencies of the non-polar (mTO) and polar (LO, PP) $p$-type modes. The $C_p(x)$ terms act as supplementary weighting factors on the individual $p$-related Raman intensities.

The Raman spectra are recorded at a quasi-fixed scattering angle $\theta = \widehat{(\vec{k_i},\vec{k_s})}$ between the adjustable-incident and fixed-scattered wavevectors inside the crystal. The $q$-dependence of the RCS is transformed into a proper $\theta$-dependence, more appropriate for comparison with experiment, via the scalar form of the wavevector conservation $\vec{q} = \vec{k}_i - \vec{k}_s$, that governs the $\vec{q}$-transfer to the crystal. The subscripts refer to the incident and scattered lights. The accurate $\theta$ value per forward-RS spectrum was derived from the criteria that the experimentally probed $q(\omega)$ dispersion (referred to as the Raman scan line in Fig. 3), would cross the zb-$Zn_{1-x}Mn_xTe$ branches $q(\omega)$ PP dispersions exactly at the experimentally observed PP frequencies. In practice, the mTO modes are visible on top of the PP modes in the forward-RS spectra. This is because the incident laser generates a (spurious) backward-RS (mTO-like) signal after reflection at the top crystal face besides the (regular) forward-RS (PP-like) signal generated by the laser beam on its way forth to the top crystal face.

Despite the small transfer of (far-infrared) energy to the crystal by Raman scattering (operated in the visible), *i.e.*, merely of the order of one percent, meaning that $k_i \sim k_s$, $q_{min} = |k_i \sim k_s| = c^{-1} \times |n(\omega_i) \times \omega_i - n(\omega_s) \times \omega_s|$ achieved in perfect forward-RS ($\theta = 0°$) can never be zero in a cubic crystal (at $\omega_i \neq \omega_s$). This is because the refractive index of cubic semiconductors reduces with energy below the optical band gap[2] where the ("transmission-like") forward-RS is done. So, the difference in energy between the incident and scattered lights is emphasized by the corresponding difference in refractive indices of the crystal, whether using the Stokes ($\omega_i > \omega_s$) or anti-Stokes ($\omega_i < \omega_s$) scattering geometries. The current $n(\omega)$ dispersions are derived from the reported data in the literature, measured by spectroscopic ellipsometry with zb-$Zn_{1-x}Mn_xTe$ epitaxial layers[15]. Direct ellipsometry measurements on our single crystals are prevented by possible crystal inhomogeneity and twins at the surface scale of few mm$^2$ probed by ellipsometry, evidenced in the large crystals studied by INS.

## Data availability

The data that support the findings of this study are available from the corresponding author upon reasonable request.

## Acknowledgements

We acknowledge assistance from the IN22 beamline staff of Institut Laüe Langevin (ILL, Proposal CRG-3343, A.E., N.B., F.B., P.B., M.M. & O.P. (2026). Effect of size versus mass disorder on phonon behavior in disordered $Zn_{1-x}Mn_xTe$ and $Mg_{1-x}Mn_xTe$ semiconductor alloys), from the PSICHÉ beamline staff of synchrotron SOLEIL (Proposal 20210410 – BAG for PSICHÉ beamline led by Y.L.G., sub-project HP-ZnMnTe, A.P., M.M. & O.P.), from the IJL core facility (Université de Lorraine – http://ijl.univ-lorraine.fr/recherche/centres-de-competences/rayons-x-et-spectroscopie-moessbauer-x-gamma) for the X-ray diffraction measurements at ambient pressure, from the Ellipsometry platform of LCP-A2MC – https://lcp-a2mc.univ-lorraine.fr/plateformes/ellipsometrie), and from Mathieu Hebmann for technical assistance in Raman scattering. Calculations have been done using the facilities of the EXPLOR HPC Center of the *Université de Lorraine* (project 2019CPMXX0918). This work was funded by the French PIA project *Lorraine Université d'Excellence* (LUE), part of the France 2030 Program, reference ANR-15-IDEX-04-LUE, within the ViSA – IRP «**Vi**brations of **S**emiconductor **A**lloys – **I**nternational **R**esearch **P**artnership» wall-less associated international laboratory (2024 – 2028), co-funded by the Excellence Initiative – Research University program at Nicolaus Copernicus University in Toruń (Poland). We further acknowledge financial support from the 2FDN – *Fédération Française de la Diffusion Neutronique*.

## Author contributions

High-quality $Zn_{1-x}Mn_xTe$ mixed crystals with zincblende structure were grown using the Bridgman method by T.N. & K.S. Powder X-ray diffraction measurements at ambient pressure were done in laboratory conditions by S.D., completed by high-pressure ones at the PSICHÉ beamline of synchrotron SOLEIL by M.M., A.P. & O.P. with the assistance of Y.L.G., N.G. & J.P.I. The high-pressure data were subsequently analyzed by M.M. The alloy composition $x$ was determined by S.M. applying the inductively coupled plasma-optical emission spectroscopy method to powders. Prospective ellipsometry measurements in the visible on selected crystal pieces were recorded by A.E & L.B. on crystal faces polished to optical quality by P.F. & A.M. *Ab initio* mTO-phonon calculations at $x \rightarrow 0$, useful to guide experiments, were done by A.V.P. using the SIESTA code. Inelastic neutron scattering measurements on large crystal pieces were done at the IN22 beamline of Institut Laue Langevin by A.E., M.M. & O.P. with the assistance of P.B., F.B. & N.B. Backward/forward Raman spectra onto/through tiny single crystals were recorded by M.M., T.N., S.T., T.T. & O.P. with the technical assistance of P.F. Analytical calculations of the (phonon-polariton, mTO, LO) Raman cross section (RCS) within the linear dielectric formalism done by O.P. & M.M. were implemented by M.M. using the MATLAB programming language. M.M. made the figures. O.P. wrote the manuscript, with contributions from all co-authors.

## Competing interests

The authors declare no competing interests.

## Additional information

**Supplementary information** The online version contains supplementary material concerning additional experimental and theoretical data further supporting the discussion of the lattice structure (X-ray diffraction) and lattice dynamics (Raman scattering) of zb-$Zn_{1-x}Mn_xTe$, available at https://...

**Correspondence** and requests for materials should be addressed to Olivier Pagès.

## Affiliations

Université de Lorraine, LCP-A2MC, F-57000 Metz, France
M. Mina, T. Ngulezhu, S. Thiam, A. V. Postnikov, T. Tite, A. En Naciri, L. Broch, P. Franchetti & O. Pagès
Laboratoire Léon Brillouin, CEA-CNRS, Saclay, France
A. Elmahjoubi & P. Bourges
Institute of Physics, Faculty of Physics, Astronomy and Informatics, Nicolaus Copernicus University in Toruń, ul. Grudziądzka 5, 87-100 Toruń, Poland
T. Ngulezhu & K. Strzałkowski
Université Grenoble Alpes, CEA, IRIG, MEM, MDN, F-38000 Grenoble, France
F. Bourdarot & N. Biniskos
Institut de Minéralogie, de Physique des Matériaux et de Cosmochimie, Sorbonne Université — UMR CNRS 7590, F-75005 Paris, France
A. Polian & Y. Le Godec
Synchrotron SOLEIL, L'Orme des Merisiers Saint-Aubin, BP 48 F-91192 Gif-sur-Yvette Cedex, France
Y. Le Godec, N. Guignot & J.-P. Itié
Université de Lorraine, CNRS, IJL, F-57000, Metz, France
S. Diliberto & S. Michel
Université de Lorraine, CentraleSupélec, LMOPS, F-57000 Metz, France
A. Maillard

## Funding

ViSA-IRP project *referenced ANR-15-IDEX-04-LUEwithin the French PIA project «Lorraine Université d'Excellence», part of the France 2030 Program.*
Synchrotron SOLEIL (Proposal 20210410) within BAG for PSICHÉ beamline (led by Y.L.G.).
2FDN – *Fédération Française de la Diffusion Neutronique* within Proposal CRG-3343 (led by A.E.) on the IN22 beamline of Institut Laüe Langevin.
Mesocenter of calculation EXPLOR at the University of Lorraine within Project project 2019CPMXX0918 (led by A.V.P.)

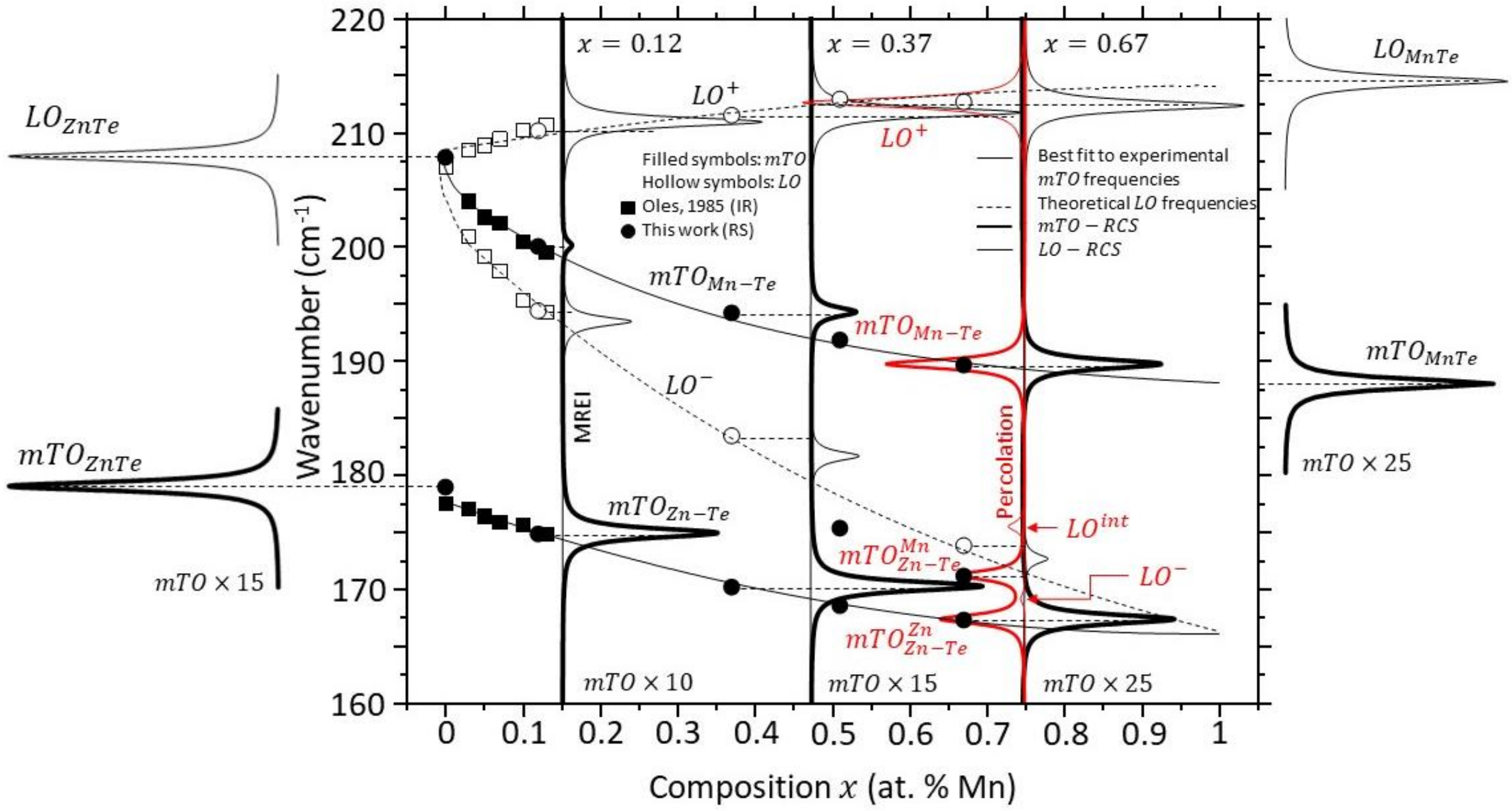

**Figure 1. Overview of zb-$Zn_{1-x}Mn_xTe$ ($x \leq 0.7$) optical (mTO, LO) modes at ambient conditions.** Existing mTO (filled symbols) and LO (empty symbols) frequencies, and current Raman data (circles) are indicated. Solid lines: quadratic mTO-frequency vs. $x$ MREI-like (two-mode) descriptions. Dotted lines: corresponding LO-frequency vs. $x$ variations derived from the LO-RCS. Thick curves: mTO MREI- (2-mode, right-oriented black curves/notations) and percolation-type (3-mode, left-oriented red curves/notations) RCS at the experimentally studied $x$ values. Thin curves: Corresponding LO MREI- (right-oriented black curves) and percolation-type (left-oriented red curves) RCS. Outside main panel: mTO (thick curves) and LO (thin curves) RCS of the parent compounds. A small phonon damping (1 cm$^{-1}$) is used throughout to distinguish small and/or neighboring features.

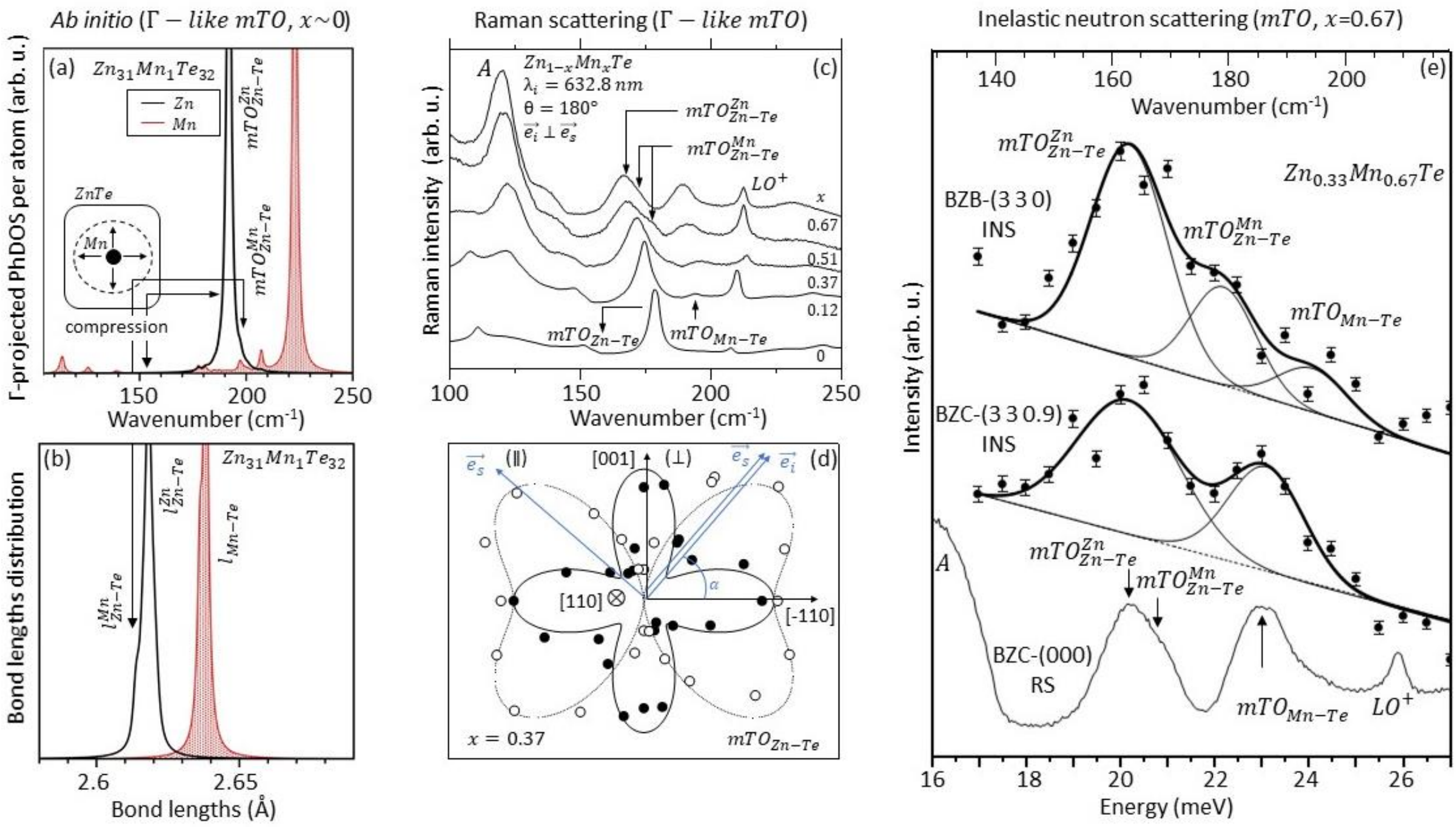

**Figure 2. Zb-Zn$_{1-x}$Mn$_x$Te mTO regime. (a)** *Ab initio* (SIESTA code) Zn-Te (curve) and Mn-Te (shaded curve) mTO $\Gamma$-projected PhDOS of a large (2×2×2) zb-$Zn_{31}Mn_1Te_{32}$ supercell ($x\sim0$). Mn creates a local compression, as sketched out (diverging arrows), impacting the phonon frequencies and bond lengths (trans-panel arrows). **(b)** Corresponding bond lengths ($l$) distribution. **(c)** Crossed-polarized pure-mTO Zn$_{1-x}$Mn$_x$Te backward-RS spectra recorded with the incident laser excitation $\lambda_i$ =632.8 nm across the zb-domain ($x \leq$0.7). **(d)** $Zn_{0.63}Mn_{0.37}Te$ $mTO^{Zn}_{Zn-Te}$ Raman intensity as a function of the azimuth angle $\alpha$ in crossed (filled symbols, ⊥) and parallel (hollow symbols, ∥) polarizations of the incident ($\vec{e}_i$) and scattered ($\vec{e}_s$) lights. The theoretical Greekcross- (solid line) and butterfly-like (dotted line) motifs are superimposed, for comparison. **(e)** $Zn_{0.33}Mn_{0.67}Te$ mTO INS spectra recorded along [00$\bar{1}$] near the BZC-(331) (at 10% of the BZsize, central curve) and near the BZB-(330) (at 90% of the BZsize, top curve). The neutron data (symbols) are fitted (thick curves) by two (∼BZC) and three (BZB) Gaussian functions (thin curves) using oblique baselines. The mTO backward-RS spectrum near the BZC-(000) (at 1% of the BZsize, bottom curve) is added for reference purpose. $A$ denotes acoustic modes.

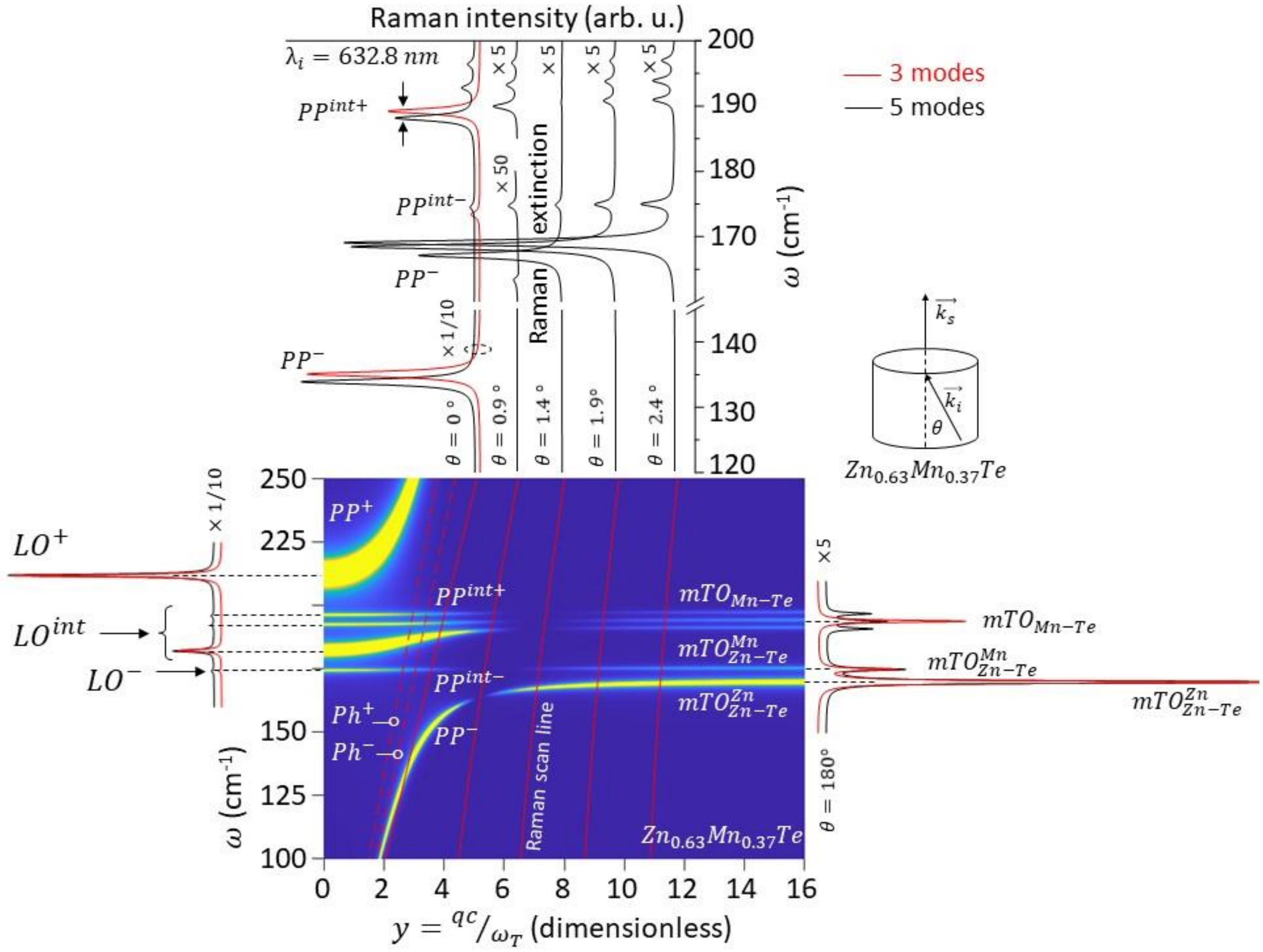

**Figure 3. Zb-Zn$_{0.63}$Mn$_{0.37}$Te phonon-polariton Raman cross section (PP-RCS). Main panel.** Zn$_{0.63}$Mn$_{0.37}$Te $q$-dependent PP-RCS in a generic percolation-type three-mode $\{2 \times (Zn - Te), 1 \times (Mn - Te)\}$ description including a triple fine structure of the $mTO_{Mn-Te}$ oscillator (see text for detail). The PP dispersion (frequency aspect, curves) and Raman efficiency (intensity aspect, thickness of curves) are indicated. $Ph^{\pm}$ denote the phonon asymptotes (dotted/red lines, methods). Raman scan lines at various scattering angles $\theta$ for the incident laser excitation $\lambda_i$ =632.8 nm are indicated (oblique/red curves). **Right.** Native mTO-RCS of the PPs at large $q$ in the crude three-mode (thick/red curve) and refined "three-mode + fine structure" (thin/black curve) descriptions. **Left.** Corresponding asymptotic LO-RCS limits of the PPs at $q$=0 (using the same code). **Top.** Zn$_{0.63}$Mn$_{0.37}$Te $\theta$-dependent PP-RCS throughout the PP regime in the refined "three-mode + fine structure" description (thin curves) calculated for the 632.8 nm incident laser excitation. $\omega_T$ and $c$ used to define the dimensionless $y$ parameter from $q$ are the mTO frequency of pristine ZnTe and the speed of light in vacuum, respectively. A small phonon damping (1 cm$^{-1}$) is used throughout to distinguish small and/or neighboring features.

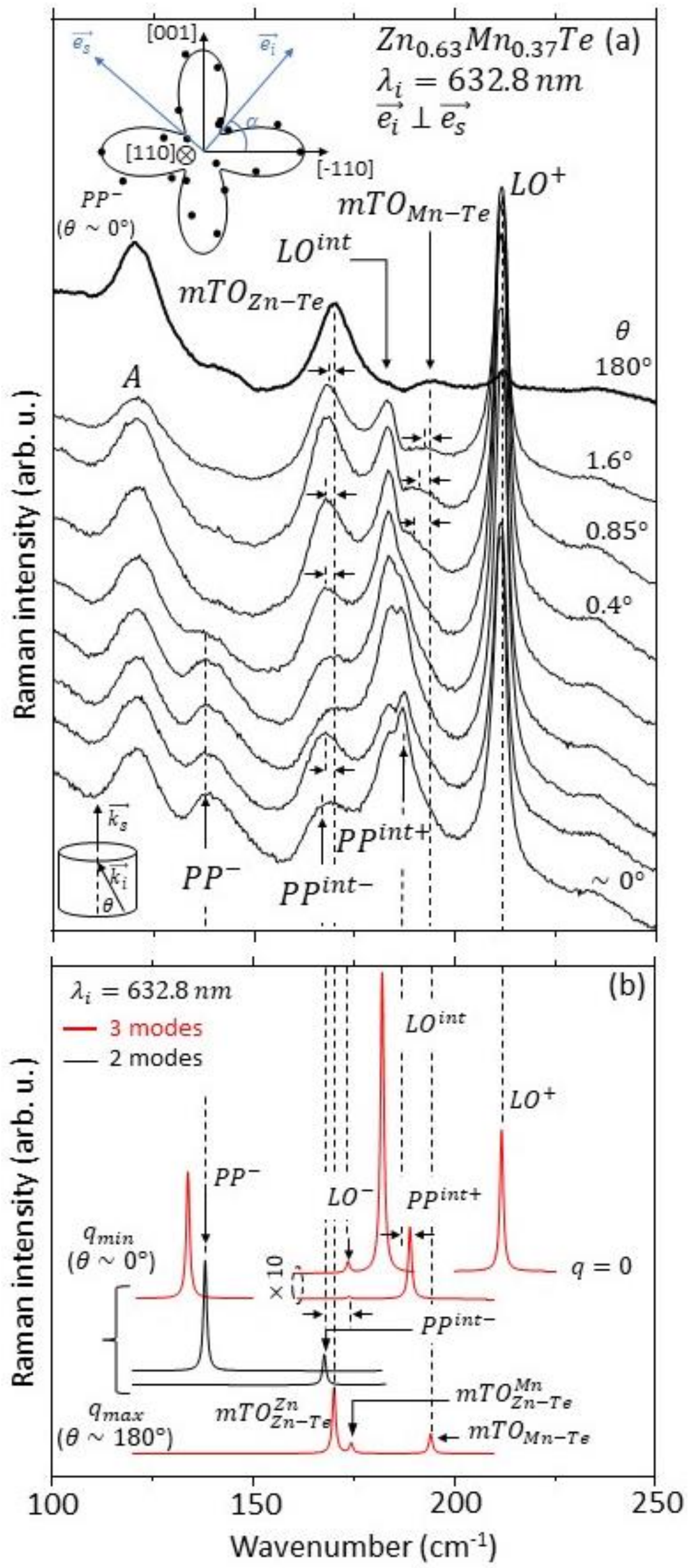

**Figure 4. Zb-$Zn_{0.67}Mn_{0.33}Te$ forward-RS spectra across the phonon-polariton regime. (a)** Cross-polarized $Zn_{0.67}Mn_{0.33}Te$ forward-RS PP spectra depending on the scattering angle $\theta$ inside the crystal, as sketched out, taken with the incident laser excitation $\lambda_i$ =632.8 nm. The corresponding backward-RS spectra taken at the same crystal spot in the mTO regime of the PPs is added (top/thick curve), for reference purpose. **Inset.** $Zn_{0.63}Mn_{0.37}Te$ $PP^-$ Raman intensity at $\theta_{min}$~0° in its dependence on the azimuth angle $\alpha$ (Fig. 2) in crossed (filled symbols) polarizations ($\vec{e}_i \perp \vec{e}_s$). The raw data are shown in Fig. S4. The theoretical Greekcross-like (solid line) motif is superimposed, for comparison. **(b)** Nominal percolation-type 3-mode $\{2 \times (Zn - Te), 1 \times (Mn - Te)\}$ (red curves) PP-RCS at $q_{min}$ ($\theta$=0°) for the 632.8 nm laser line and corresponding asymptotic LO ($q$=0) and mTO ($q_{max}$) RCS. Truncated 2-mode $\{1 \times (Zn - Te), 1 \times (Mn - Te)\}$ descriptions are used to generate the uncoupled $PP^-$ and $PP^{int+}$ RCS (black curves) separately due to $mTO^{Zn}_{Zn-Te}$ and $mTO^{Mn}_{Zn-Te}$, respectively, at $q_{min}$. Dashed guidelines are used to facilitate the comparison between the experimental (a) and theoretical (b) Raman frequencies. A small phonon damping (1 cm$^{-1}$) is used throughout to distinguish small and/or neighboring features. $A$ denotes acoustic modes.

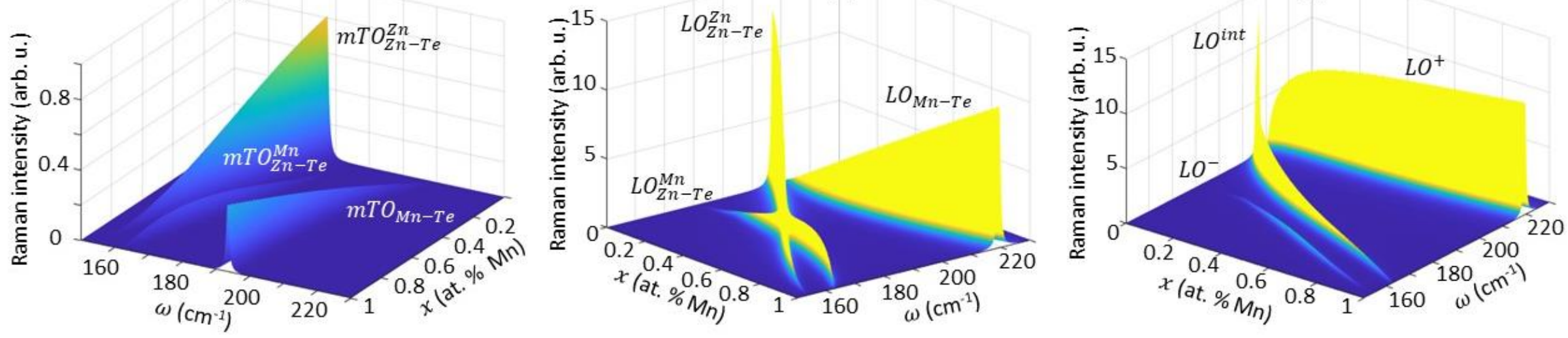

**Figure 5. Percolation-type 3-oscillator $\{2 \times (Zn - Te), 1 \times (Mn - Te)\}$ $Zn_{1-x}Mn_xTe$ RCS in their $x$ - dependence. (a)** Mechanical-TO (mTO) RCS **(b)** Uncoupled-LO RCS. **(c)** Coupled-LO RCS. A small phonon damping (1 $cm^{-1}$) is used throughout to distinguish small and/or neighboring features.

## SUPPLEMENTARY INFORMATION

## Cubic-$Zn_{1-x}Mn_xTe$ optical modes across the phonon and phonon-polariton regimes

M. Mina,[1] A. Elmahjoubi,[2] T. Ngulezhu,[1,3] S. Thiam,[1] A. V. Postnikov,[1] T. Tite,[1] P. Bourges,[2] F. Bourdarot,[4] N. Biniskos,[4] A. Polian,[5] Y. Le Godec,[5,6] N. Guignot,[6] J.-P. Itié,[6] S. Diliberto,[7] S. Michel,[7] A. Maillard,[8] A. En Naciri,[1] L. Broch,[1] P. Franchetti,[1] K. Strzałkowski[3] and O. Pagès[1,a]

[1] Université de Lorraine, LCP-A2MC, F-57000 Metz, France
[2] Laboratoire Léon Brillouin, CEA-CNRS, Saclay, France
[3] Institute of Physics, Faculty of Physics, Astronomy and Informatics, Nicolaus Copernicus University in Toruń, ul. Grudziądzka 5, 87-100 Toruń, Poland
[4] Université Grenoble Alpes, CEA, IRIG, MEM, MDN, F-38000 Grenoble, France
[5] Institut de Minéralogie, de Physique des Matériaux et de Cosmochimie, Sorbonne Université — UMR CNRS 7590, F-75005 Paris, France
[6] Synchrotron SOLEIL, L'Orme des Merisiers Saint-Aubin, BP 48 F-91192 Gif-sur-Yvette Cedex, France
[7] Université de Lorraine, CNRS, IJL, F-57000, Metz, France
[8] Université de Lorraine, CentraleSupélec, LMOPS, F-57000 Metz, France

([a] Author to whom correspondence should be addressed: olivier.pages@univ-lorraine.fr)

Supplementary information (prefixed S) reports on additional experimental and theoretical data further supporting the discussion of the lattice structure (Sec. SI) and dynamics (Sec. SII) of $Zn_{1-x}Mn_xTe$ taken in the zb-phase, in relation to X-ray diffraction (XRD) and Raman scattering (RS), respectively. In the latter case both experimental and theoretical data are reported, concerned with backward/mTO-like and forward/PP-like Raman signals, and with the intra-bond LO-coupling within the Zn-Te dual optical mode of zb-$Zn_{1-x}Mn_xTe$, respectively.

### SI. $Zn_{1-x}Mn_xTe$ lattice structure

Raw $Zn_{1-x}Mn_xTe$ XRDiffractograms obtained at minor ($x$=0.12) and large ($x$=0.37) alloying at the PSICHÉ beamline of SOLEIL synchrotron using the 0.3738 Å radiation at nearly ambient pressure by inserting finely ground powders inside a diamond anvil cell is displayed in Fig. S1. Au and Ne are used for pressure calibration and as the pressure transmitting medium, respectively. An additional powder XRDiffractogram related to large Mn content ($x$=0.67) was separately measured at ambient pressure in laboratory conditions using the 1.541 Å Cu K$\alpha$ line (not shown). Sharp diffraction peaks, attesting a high crystal quality, indicate that all studied samples crystallize in the same zincblende (zb) structure as ZnTe. The individual XRDiffraction peaks in Fig. S1 are labelled accordingly via their (hkl) Miller's indices.

### SII. $Zn_{1-x}Mn_xTe$ lattice dynamics

1. Backward-RS

Fig. S2 displays the full series of cross-polarized backward-RS spectra taken at the same sample spot with the 632.8 nm laser line across the zb-phase of $Zn_{1-x}Mn_xTe$ ($x \leq$0.7) depending on the azimuth angle $\alpha$ between the incident polarization ($\vec{e}_i$) and the $[\bar{1}10]$ crystal axis taken as an arbitrary reference, step-increased by 20° until completing a full revolution (over 360°) at the sample surface. The reported symmetry analysis at $x$=0.37 in the inset of Fig. 2 refers to the $\alpha$-dependence of the normalized (to the largest signal) Raman intensity (area) of the main ZnTe-like peak (modeled as a Lorentzian/Gaussian function) in crossed (⊥, filled symbols) and parallel (∥, hollow symbols) polarizations of the incident ($\vec{e}_i$) and scattered ($\vec{e}_s$) lights. The Zn-Te and Mn-Te Raman intensities exhibit the same $\alpha$-dependence corresponding to the theoretical quatrefoil Greekcross-like (⊥) and bilobed butterfly-like (∥) motifs axed $[\bar{1}10]$ and $[001]$, as ideally expected. Similar motifs are observed at $x$=0.12 (Fig. S2g) and $x$=0.67 (Fig. S2h). So, the lower/main Zn-Te and upper/minor Mn-Te Raman peaks consist of genuine mTO modes. The raw cross-top/parallel-bottom polarized Raman spectra collected at all studied $x$ values (0.12, 0.37, 0.67) are displayed in Fig.S2. In each panel, the extreme (top and bottom) curves, referring to the same $\alpha$ addressed at the start and at the end of the revolution, help to appreciate the stability of the laser source across the full $\alpha$-revolution.

2. Forward-RS

Fig. S3 displays a selection of cross-polarized forward-RS spectra obtained at various scattering angles $\theta$ with zb-$Zn_{0.33}Mn_{0.67}Te$ using the 632.8 nm laser line at quasi normal incidence/detection across parallel (110)-oriented crystals faces obtained by cleavage. The specified $\theta$ per spectrum are estimated so that the corresponding $\theta$-dependent Raman scan line crosses the $q$-dependent zb- zb-$Zn_{0.33}Mn_{0.67}Te$ PP-RCS (comparable to that reported for $x$=0.37 in Fig. 3) exactly at the experimentally observed $PP^{-}$ (small $\theta$) or $PP^{int+}$ (large $\theta$) frequencies. By decreasing $\theta$, the PPs progressively soften from their native Zn-Te and Mg-Te mTO modes. The $PP^{-}$ mode hardly departs from its native $mTO_{Zn-Te}^{Zn}$ mode as long as $\theta \geq$0.95° (Fig. S3) and then suddenly emerges as a distinct feature downshifted (frequency aspect) far below (by as much as ~50 cm$^{-1}$) the $mTO_{Zn-Te}^{Zn}$ mode at $\theta \leq$0.85° (Fig. S3). No intermediate stage is observed. This is consistent with $PP^{-}$ suffering a massive Raman extinction (intensity aspect) when it gets massively photon-like on crossing the photon asymptotes, which precisely occurs at the critical $\theta$ value $\theta_c$~0.90° (Fig. 3). Remarkably, $\theta_c$ is also critical with respect to the $PP^{int+}$ mode, subject to a broad↔sharp transition (linewidth aspect) at this limit (Fig. S3). This cannot be merely fortuitous, proving on an experimental basis that the latter transition can be traced back to the crossing of the photon-like asymptotes. The $\theta_c$-coincidence affecting the $PP^{-}$ and $PP^{int+}$ modes is presumably intrinsic to zb-$Zn_{1-x}Mn_xTe$ alloying since it is independently observed with zb-$Zn_{0.63}Mn_{0.37}Te$ (Fig. 4).

Fig. S4 shows the raw forward-RS cross-polarized spectra at minimum scattering angle ($\theta$~0°) behind the reported $PP^{-}$ symmetry analysis at $x$=0.37 (Fig. 4, inset) and $x$=0.67 (Fig. S3, inset) using the 632.8 nm laser line. Similar spectra for $x$=0.12 complete the series. At a given $x$, the forward-RS PP-symmetry analysis is done at the same sample spot as the backward-RS mTO-symmetry analysis in Fig. S2. The specified $\alpha$ values do correspond. The $PP^{-}$ Raman intensity in its $\alpha$-dependence exhibits the same Greekcross-like motif (inset of Fig. 4, filled symbols) as that of the native $mTO_{Zn-Te}^{Zn-Te}$ mode (inset of Fig. 2, filled symbols) attesting that they both consist of genuine TO modes. More generally, all three $PP^{-}$, $PP^{int-}$ and $PP^{int+}$ follow the same Raman selection rules depending on $\alpha$ (as apparent at any $x$ in Fig. S4) attesting their identical (PP) nature.

3. Inter-bond LO-coupling – simplified two-oscillator MREI-like description

In the main text the inter-bond LO-coupling in $Zn_{1-x}Mn_xTe$ with zb-phase artificially expanded across the entire composition domain is treated in full by using the nominal three-oscillator percolation-type approach, referring to the native $\{mTO_{Zn-Te}^{Zn}, mTO_{Zn-Te}^{Mn}, mTO_{Mn-Te}\}$ modes behind the LO modes. The corresponding oscillator fractions $f_p(x)$ are listed in methods. A simplified MREI-like two-oscillator $\{mTO_{Zn-Te}, mTO_{Mn-Te}\}$ description is achieved by merging the percolation-type $mTO_{Mn-Te}$–$mTO_{Zn-Te}^{Zn}$ duo into a unique MREI-like $mTO_{Zn-Te}$ oscillator with fraction $f_{Zn-Te}(x) = 1 - x$. The remaining Mn-Te oscillator, already of the MREI-type, is represented in the crystal with fraction $f_{Mn-Te}(x) = x$. The corresponding two-mode uncoupled-$\{LO_{Zn-Te}, LO_{Mn-Te}\}$ and coupled-$\{LO^{-}, LO^{+}\}$ RCS are displayed in Fig. S5, using a minimum phonon damping (1 cm$^{-1}$). The $E_L$-coupling is manifested by a strong LO repulsion at the resonance ($x$~0) due to a massive $LO_{Zn-Te} \rightarrow LO_{Mn-Te}$ transfer of oscillator strength, reflected in the $LO^{-}$ (minor) and $LO^{+}$ (dominant) Raman intensities (Fig.

S5-top, thickness of curves), in line with observed trends within the three-oscillator scheme (Fig. 5). The simplified two-oscillator scheme is, notably useful to compare on equal footing zb-$Zn_{1-x}Mn_xTe$ with zb-$Cd_{1-x}Zn_xTe$ and zb-$Ga_{1-x}In_xAs$, earlier treated in the same (MREI-like) way[29]. The comparison involves the dimensionless parameter $\tau$ (Fig. S5c), introduced in the main text, representing the amount of Zn-Te/Mn-Te mixing per coupled-LO mode depending on $x$. The full treatment, outlined in the main text offering the access to $\tau$, is detailed in Ref. 29, referring specifically to Eqs. 3-to-5 therein. Remarkably, near the resonance ($x \sim 0$), a higher mixing is achieved for zb-$Zn_{1-x}Mn_xTe$ ($\tau_{max}$~0.5 at $x \sim 0.10$) than for $Cd_{1-x}Zn_xTe$ and zb-$Ga_{1-x}In_xAs$ ($\tau_{max}$~0.4 at $x \sim 0.05$) – Ref. 29 (Fig. 7d therein).

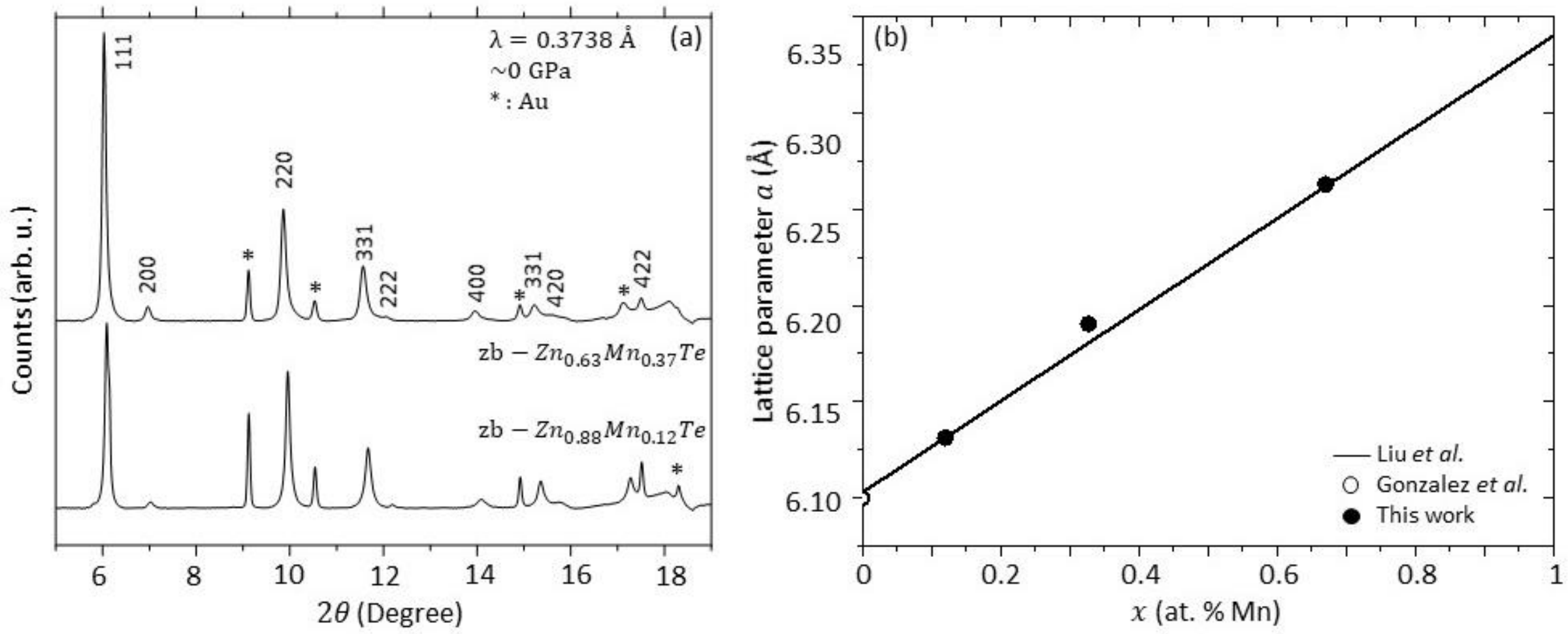

**Figure S1. $Zn_{1-x}Mn_xTe$ crystal structure. (a)** Powder X-ray diffractograms recorded at nearly ambient pressure on the PSICHÉ beamline of synchrotron SOLEIL with the 0.3738 Å radiation at $x$=0.12 and 0.37. The diffraction lines due to Au used for pressure calibration are indicated. The individual peaks are labeled using the (hkl) Miller indices of the corresponding diffraction planes in the zincblende-zb structure. **(b)** $a(x)$ dependence (symbols) derived from the lattice parameter $a$ and the composition $x$ independently measured in laboratory by powder X-ray diffraction with the Cu K-alpha radiation at ambient pressure and by applying the inductively coupled plasma-optical emission spectroscopy method to powders, respectively. The linear $a(x)$ variation found by Liu et al.[15] using epitaxial layers is added for comparison.

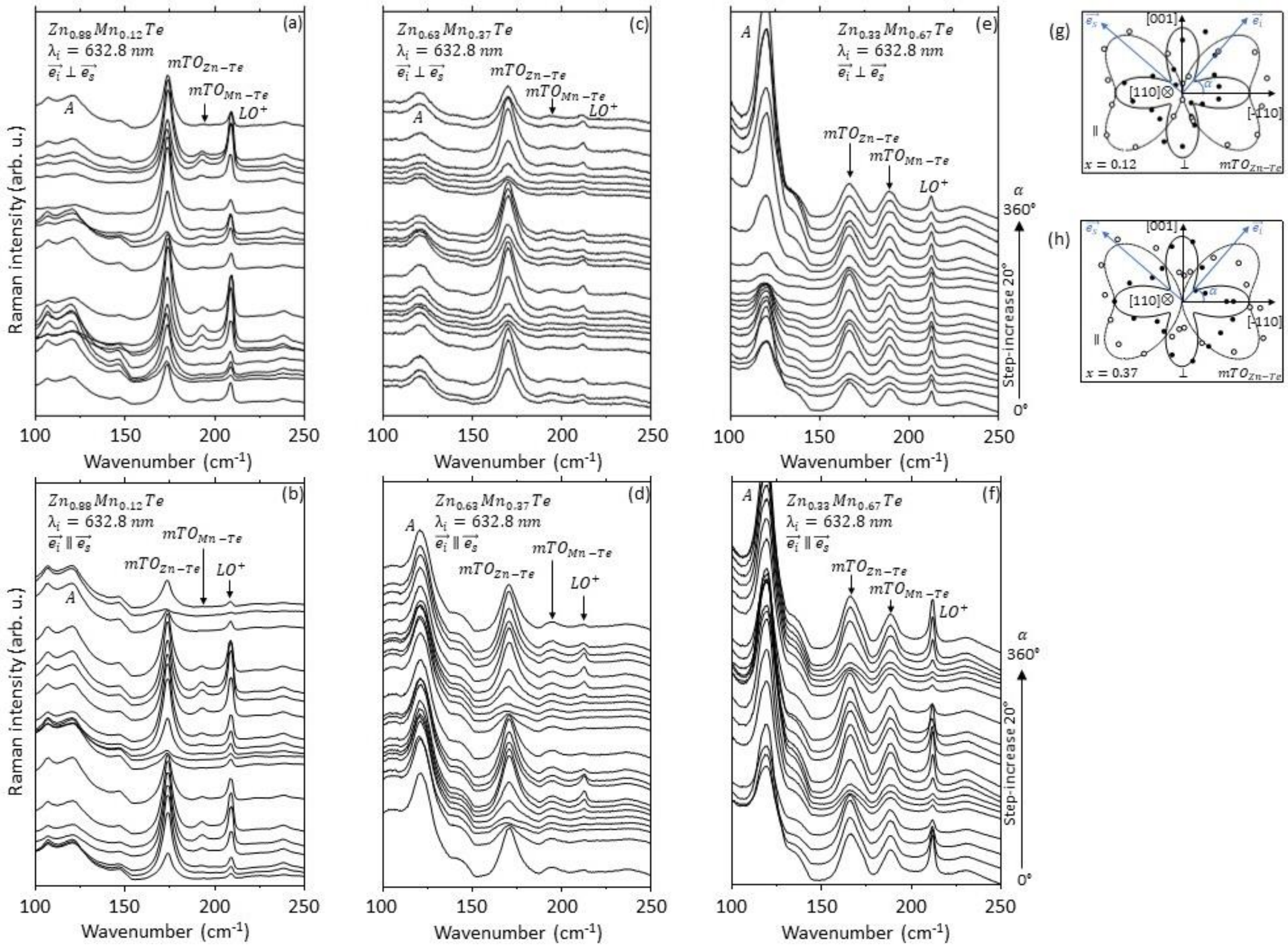

**Figure S2.** Polarized $Zn_{1-x}Mn_xTe$ mTO backward-RS spectra taken with the incident laser excitation $\lambda_i$ =632.8 nm. Cross (top, ⊥) and parallel (down, ǁ) polarized $Zn_{1-x}Mn_xTe$ mTO backward-RS spectra excited at the same sample spot of (110)-cleaved crystal faces depending on the azimuth angle $\alpha$ between the incident polarization ($\vec{e}_i$) and the $[\bar{1}10]$ crystal axis across the zb-phase, as sketched out (right). **(a/b)** $x$=0.12. **(c/d)** $x$=0.37. **(e/f)** $x$=0.67. **(g/h).** Corresponding experimental (symbols) vs. theoretical (curve) Greekcross- (⊥) and butterfly-like (ǁ) $\alpha$-variations of the Zn-Te Raman intensity at small ($x$=0.12, g) and large ($x$=0.67, h) alloying. Corresponding data at moderate alloying ($x$=0.67) are shown in Fig. 2d. $A$ denotes acoustic modes.

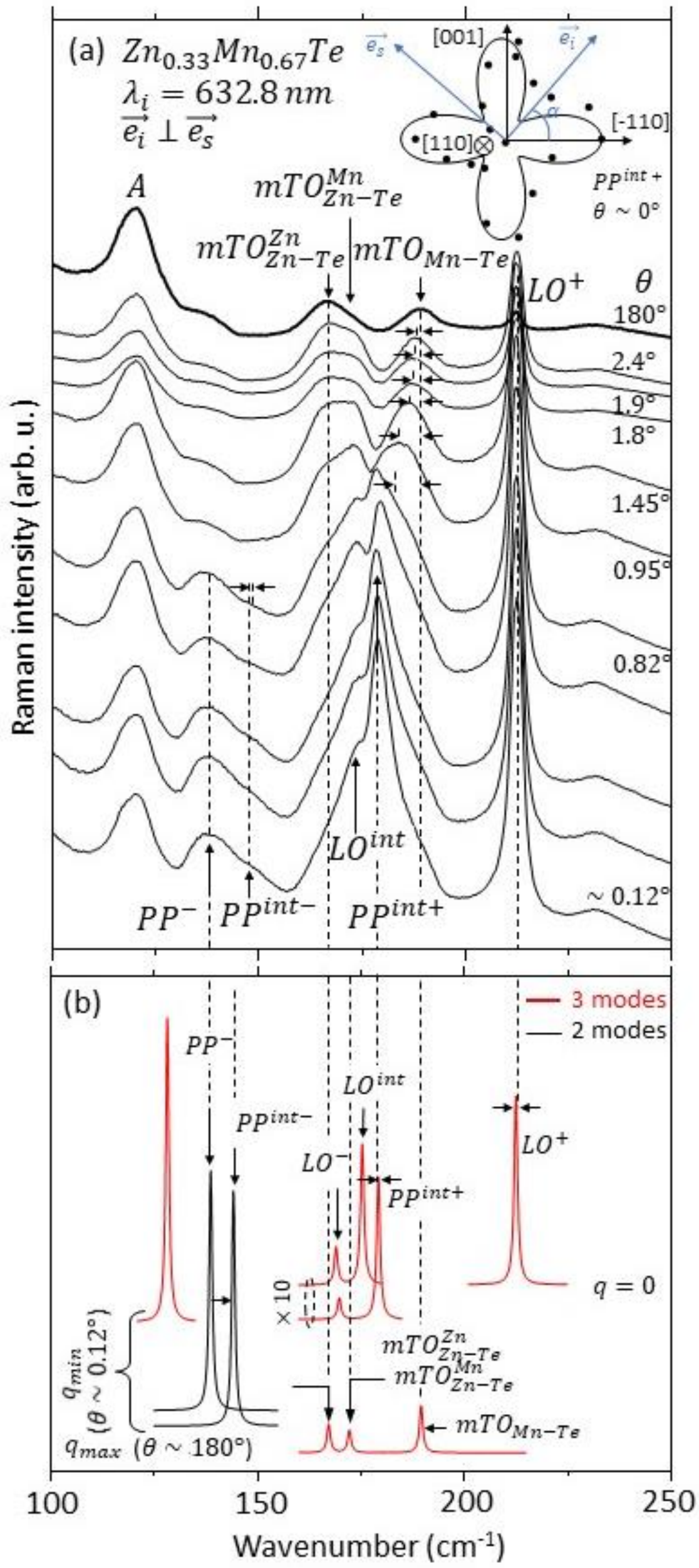

**Figure S3.** $Zn_{0.33}Mn_{0.67}Te$ forward-RS spectra across the phonon-polariton regime. **(a)** Cross-polarized $Zn_{0.33}Mn_{0.67}Te$ forward-RS PP spectra depending on the scattering angle $\theta$ inside the crystal, taken with the incident laser excitation $\lambda_i$ =632.8 nm. The corresponding backward-RS spectra taken at the same crystal spot in the native mTO regime of the PPs is added (top/thick curve), for reference purpose. **Inset.** $Zn_{0.63}Mn_{0.67}Te$ $PP^{int+}$ Raman intensity at $\theta_{min}$~0° as a function of the azimuth angle $\alpha$ (as sketched out in Fig. 2) in crossed (filled symbols) polarizations ($\vec{e}_i \perp \vec{e}_s$). The raw data are shown in Fig. S4. The theoretical Greekcross-like (solid line) motif is superimposed, for comparison. **(b)** Nominal percolation-type 3-mode $\{2 \times (Zn - Te), 1 \times (Mn - Te)\}$ (red curves) PP-RCS at $q_{min}$ ($\theta$~12°) for the 632.8 nm laser line and corresponding asymptotic LO ($q$=0) and mTO ($q_{max}$) RCS. Truncated 2-mode $\{1 \times (Zn - Te), 1 \times (Mn - Te)\}$ descriptions are used to generate the uncoupled $PP^-$ and $PP^{int+}$ RCS (black curves) separately due to $mTO^{Zn}_{Zn-Te}$ and $mTO^{Mn}_{Zn-Te}$, respectively, at $q_{min}$ (see main text). Dashed guidelines are used to facilitate the comparison between the experimental (a) and theoretical (b) Raman frequencies. A small phonon damping (1 cm$^{-1}$) is used throughout to distinguish small and/or neighboring features. $A$ denotes acoustic modes.

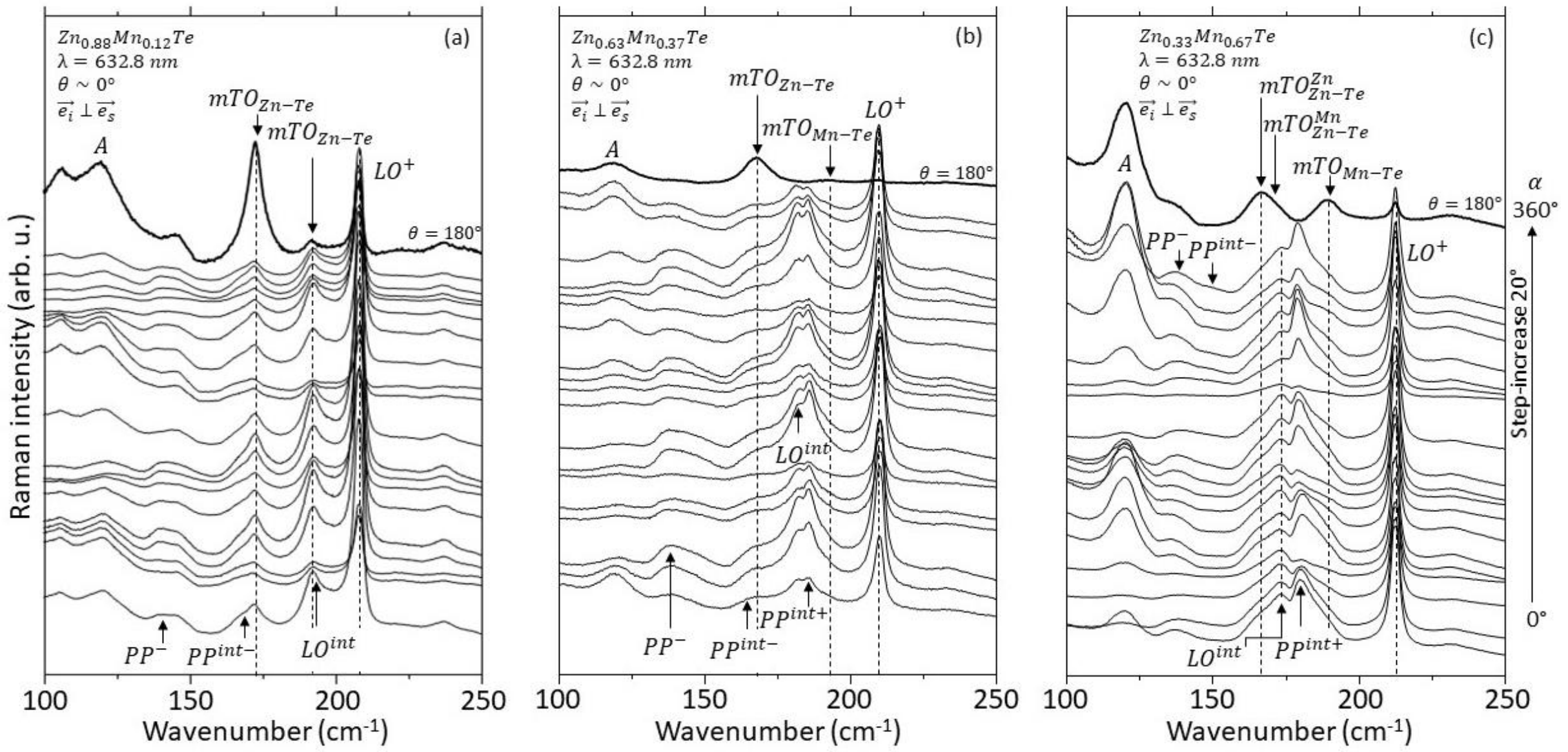

**Figure S4.** Polarized zb-$Zn_{1-x}Mn_xTe$ forward-RS spectra ($\theta\sim0°$) taken with the incident laser excitation $\lambda_i$ =632.8 nm. Cross-polarized (⊥) $Zn_{1-x}Mn_xTe$ forward-RS spectra excited at the same sample spots of (110)-cleaved crystal faces depending on the azimuth angle $\alpha$ between the incident polarization ($\vec{e}_i$) and the $[\bar{1}10]$ crystal axis across the zb-phase, as sketched out in Fig. S2g. **(a)** $x$=0.12. **(b)** $x$=0.37. **(c)** $x$=0.67. $A$ denotes acoustic modes.

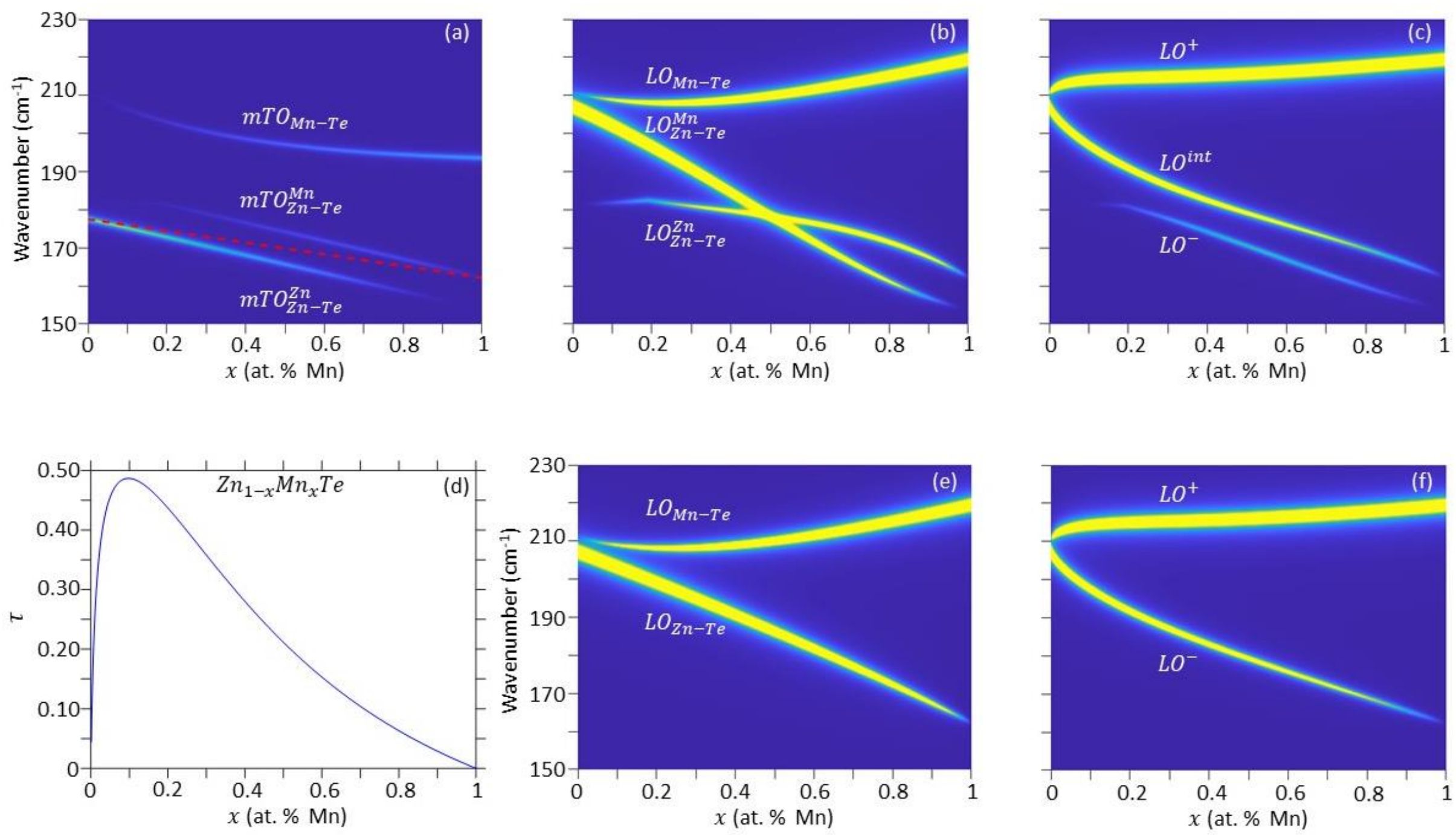

**Figure S5.** Theoretical LO-coupling in zb-$Zn_{1-x}Mn_xTe$ ($0\leq x \leq 1$) – Percolation vs. MREI approaches. **Top.** Three-mode percolation model / frontal-2D views (the corresponding tilted-3D views are displayed in Fig. 5). **(a)** Mechanical-TO (mTO) RCS. **(b)** Uncoupled-LO RCS. **(c)** Coupled-LO RCS. **Bottom.** Two-mode MREI model. The percolation mTO Zn-Te duo is grasped into a unique MREI oscillator carrying the whole available Zn-Te oscillator strength – dotted/red line in panel (a). **(d)** Amount of coupling/mixing $\tau$. **(e)** Uncoupled-LO RCS. **(f)** Coupled-LO RCS. A small phonon damping (1 $cm^{-1}$) is used throughout to distinguish small and/or neighboring features.